\documentclass[epj,final]{svjour}
\usepackage{amsmath,eucal}
\usepackage{graphics}

\newcommand{\as}{\alpha_{\mathrm{s}}}
\newcommand{\bare}{{(0)}}
\newcommand{\cut}{_{\rm cut}}
\newcommand{\dJ}[1]{\frac{\dif\sigma}{\dif J}{}\raisebox{1.5ex}{$#1$}}
\newcommand{\dif}{{\rm d}}
\newcommand{\dk}{\dif\kk}
\newcommand{\dug}{\,\raisebox{0.37pt}{:}\hspace{-3.2pt}=}
\newcommand{\e}{\varepsilon}
\newcommand{\esp}[1]{{\rm e}^{#1}}
\newcommand{\fz}{f^{(0)}}

\newcommand{\hz}{h^{(0)}}
\newcommand{\id}{\boldsymbol{1}}

\newcommand{\K}{{\CMcal K}}
\newcommand{\kd}{{\boldsymbol k}_2}
\newcommand{\kk}{{\boldsymbol k}}
\newcommand{\ku}{{\boldsymbol k}_1}
\newcommand{\lb}{{\boldsymbol l}}
\newcommand{\lra}{\leftrightarrow}

\newcommand{\MS}{\overline{\rm MS}}
\newcommand{\mbf}[1]{{\boldsymbol #1}}
\newcommand{\mscr}[1]{\mbox{\scriptsize{$#1$}}}
\newcommand{\N}{{\CMcal N}}
\newcommand{\Nf}{N_{\!f}}
\newcommand{\ord}{{\CMcal O}}
\newcommand{\PP}{{\CMcal P}}
\newcommand{\pa}{{\mathsf a}}
\newcommand{\pb}{{\mathsf b}}
\newcommand{\pg}{{\mathsf g}}
\newcommand{\pic}{\small}
\newcommand{\pp}{\boldsymbol p}
\newcommand{\pq}{{\mathsf q}}
\newcommand{\product}{*}
\newcommand{\qq}{{\boldsymbol q}}
\newcommand{\Sj}[1]{{\CMcal S}_J^{(#1)}}
\newcommand{\Sjt}[1]{\widetilde{\CMcal S}_J^{(#1)}}
\newcommand{\slarga}[2]{\raisebox{-#1mm}{\rule{0pt}{#2mm}}}
\newcommand{\sr}{\hat{s}}
\newcommand{\ui}{{\mathrm i}}
\newcommand{\ugd}{=\hspace{-3.2pt}\raisebox{0.37pt}{:}\hspace{3pt}}

\newcommand{\lab}[1]{\label{#1}}
\newcommand{\labe}[1]{\label{#1}}

\newcommand\jp[3]{{ J.~Phys. }{\textbf #1} (#2) #3}
\newcommand\npb[3]{{ Nucl. Phys. }{\textbf B #1} (#2) #3}
\newcommand\npbps[3]{{ Nucl. Phys. }{\textbf B} { (Proc. Suppl.)}{ \textbf #1} (#2) #3}
\newcommand\plb[3]{{ Phys. Lett. }{\textbf B #1} (#2) #3}                   
\newcommand\prd[3]{{ Phys. Rev. }{\textbf D #1} (#2) #3}            
\newcommand\zpc[3]{{ Z. Physik }{\textbf C #1} (#2) #3}             
\newcommand\sjnp[3]{{ Sov. J. Nucl. Phys. }{\textbf #1} (#2) #3}    
\newcommand\jetp[3]{{ Sov. Phys. JETP }{\textbf #1} (#2) #3}        
            
\newcommand{\hep}[1]{{\tt hep-ph/#1}}

\begin{document}
\headnote{\textup{\normalsize{DESY 02-090   \hfill ISSN 0418--9833 }}}
\title{
The NLO Jet Vertex for Mueller-Navelet and Forward Jets: the Gluon Part}
\author{
J.\ Bartels\inst{1}%
 \thanks{Supported by the TMR Network ``QCD and Deep Structure of Elementary Particles''.}
 \and
D.\ Colferai\inst{1}%
 \thanks{Supported by the Alexander von Humboldt Stiftung.}
 \and
G.P.\ Vacca\inst{2}%
}

\institute{II. Institut f\"ur Theoretische Physik; Universit\"at Hamburg,
 Luruper Chaussee 149, 22761 Hamburg, Germany
 \and
Dipartimento di Fisica, Universit\`a di Bologna and
Istituto Nazionale di Fisica Nucleare, Sezione di Bologna,\\
via Irnerio 46, 40126 Bologna, Italy
}
\date{Received: date / Revised version: date}
\abstract{In this paper we complete our calculation of the NLO jet vertex 
which is part of the cross section formulae for the production 
of Mueller Navelet jets at 
hadron hadron colliders and of forward jets in deep inelastic electron 
proton scattering.}  
 

\maketitle

\section{Introduction}

In a recent paper~\cite{BaCoVa01}, we have started the NLO calculation 
of the jet vertex which represents one of the building blocks in 
the production of Mueller-Navelet jets~\cite{MuNa87} at hadron hadron 
colliders and of 
forward jets~\cite{Mu90} in deep inelastic electron proton scattering. Both 
jet production processes are of particular interest for studying QCD
in the Regge limit or in the small-$x$ limit: they provide kinematical
environments for which the BFKL Pomeron~\cite{BFKL76} applies. Previous 
experience shows
that existing leading order calculations~\cite{2jetLL,FjetLL,ee} are not 
accurate enough to 
allow for a reliable comparison with experimental data. NLO calculations 
are available for the BFKL Pomeron~\cite{FaLi98,CaCi98}, but consistent 
next-to-leading order
analysis of data at the Tevatron, at LHC, or at HERA require the NLO 
calculations also of the jet vertex and of the photon impact factor. 
As to the jet vertex, in our previous paper~\cite{BaCoVa01} we have 
presented the first part, namely the quark-initiated jet vertex. In the 
present paper, we complete the NLO analysis of the jet vertex with the
gluon-initiated part. The NLO calculation of the photon impact factor 
is being pursued by two independent groups~\cite{BaGiQi00,FaMa99}. 
As a result of these 
combined efforts, it will be possible perform consistent NLO studies of 
BFKL predictions in hadron hadron colliders, in $ep$ deep inelastic 
scattering, and in $e^+e^-$ scattering processes.

A particular theoretical challenge in computing the jet 
vertex for the processes mentioned before is related to the special kinematics.
The processes to be analyzed is illustrated in Fig.\ref{f:jet}: the lower gluon 
emitted from the hadron $H$ scatters with the upper parton $\pq$ and produces 
the jet $J$. Because of the large transverse momentum of the jet, 
the gluon is hard and obeys the collinear factorization. In particular, 
its scale dependence is described by the DGLAP evolution equations \cite{DGLAP77}.
Above the jet, on the other hand, we require a large rapidity gap between 
the jet and the outgoing parton $\pq$: this kinematic requirement is described
by BFKL dynamics. Consequently, the jet vertex lies at the interface between
DGLAP and BFKL dynamics. As an essential result of our analysis we find 
that it is possible to separate, inside the jet vertex, the collinear infrared 
divergences that go into the parton evolution of the incoming gluon from 
the high energy gluon radiation inside the rapidity gap which belongs to 
first rung of the LO BFKL ladder. In~\cite{BaCoVa01} this was demonstrated 
for the quark-initiated vertex, and in the present paper we present the 
generalization to the gluon part. 

Our paper will be organized as follows. In order to make the reading
as convenient as possible, we first review the general framework,
which will be the same as in our previous
paper~\cite{BaCoVa01}. Sections \ref{s:virt} and \ref{s:rc} then contain
the virtual and real corrections, resp., and
in section \ref{s:jv} we combine both results.
Section \ref{s:jvi} contains, as a consistency check, a comparison of our jet
vertex with the gluon impact factor. In section \ref{s:vhe} we use our results
to define the general NLO jet production cross section.  The
concluding section contains a short summary and an outlook of future
steps.
            
\section{High energy factorization\labe{s:hef}}

\subsection{General framework\labe{s:gf}}

We describe the kinematics of the hadron ($H$) quark ($\pq$) collision in terms of light
cone coordinates
\begin{equation}\lab{lcc}
 p^\mu=(p^+,p^-,\pp)\;,\quad p^\pm\dug\frac{p^0 \pm p^3}{\sqrt2}\;,
\end{equation}
where the light-like vectors $p_H$ and $p_\pq$ form the basis of the longitudinal plane:
\begin{subequations}
\begin{align}
 \lab{coord1}
 p_H &= \left(\sqrt{\frac{s}{2}},0,\mbf{0}\right)\;,\quad s\dug(p_H+p_\pq)^2\\
 \lab{coord2}
 p_\pq &= \left(0,\sqrt{\frac{s}{2}},\mbf{0}\right)\\
 \lab{coord3}
 p_i &= E_i\left(\frac{\esp{y_i}}{\sqrt2},\frac{\esp{-y_i}}{\sqrt2},\mbf{\phi}_i\right)\;.
\end{align}
\end{subequations}
In the last equation we have introduced a parameterization for the $i$-th particle in the
final state in terms of the rapidity $y_i$ (in the $p_H+p_\pq$ center of mass frame), of
the transverse energy $E_i=|\pp_i|$ and of the azimuthal unit vector
$\mbf{\phi}_i\parallel\pp_i$.

\begin{figure}[hb!]
\centering
\resizebox{0.3\textwidth}{!}{\includegraphics{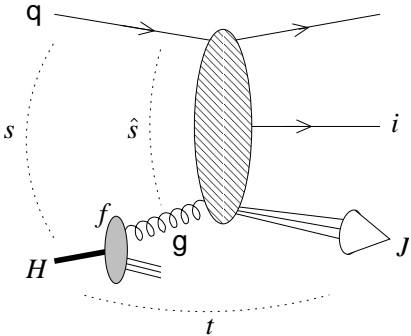}}
\caption{\labe{f:jet} Diagrammatic representation of the high energy process studied in this
  paper. $H$ is the incoming hadron providing a gluon $\pg$ with distribution density $f$;
  $\pq$ is the other incoming particle which will be taken to be a quark; $J$ denotes the
  jet produced in the forward direction (w.r.t $H$) and $i$ is the generic label for
  outgoing particles.}
\end{figure}

According to the parton model, we assume the physical cross section to be given by the
corresponding partonic cross section $\dif\hat\sigma$ (computable in perturbation theory)
convoluted with the parton distribution densities (PDF) $f_\pa$ of the partons $\pa$ inside
the hadron $H$. A jet distribution $S_J$ selects the final states contributing to the one
jet inclusive cross section that we are considering.

In terms of the jet variables --- rapidity, transverse energy and azimuthal angle ---  
the one jet inclusive cross section initiated by the gluons in hadron $H$ can be written as
\begin{equation}\lab{pSf}
 \dJ{} \dug \frac{\dif\sigma_{\pq H}}{\dif y_J \dif E_J \dif \phi_J}  
 = \int\dif x\;\dif\hat\sigma_{\pq\pg}(x)\,S_J(x)\fz_\pg(x)\;.
\end{equation}
The incoming gluon carries a fraction $x$ of the longitudinal momentum of $H$, while its
transverse motion is negligible in the high energy regime:
\begin{equation}\lab{defpa}
 p_\pg = x\,p_H =\left(x\sqrt{\frac s2},0,\mbf{0}\right)\;.
\end{equation}
In our analysis we study the partonic subprocess $\pg+\pq\to X+{\it jet}$ in the high energy limit
\begin{equation}\lab{HElimit}
 \Lambda_{\rm QCD}^2 \ll E_J^2\sim -t\;\text{ (fixed) } \ll s\to\infty
\end{equation}

\subsection{The Jet vertex at lowest order\labe{s:LO}}

\begin{figure}[b]
\centering
\resizebox{0.2\textwidth}{!}{\includegraphics{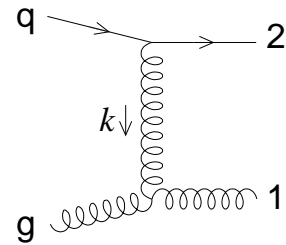}}
\caption{\label{f:gq12}Leading diagram at lowest order for gluon-quark scattering: the
  interaction occurs via gluon exchange in the $t$-channel.}
\end{figure}

In the high energy regime~(\ref{HElimit}), the lowest order (LO) contribution to the jet cross
section is dominated by gluon exchange in the $t$-channel, as shown in Fig.~\ref{f:gq12}.

The corresponding expression has been already obtained in~\cite{BaCoVa01} and is given by
\begin{equation}\lab{LOFF}
 \dJ{^{(0)}} = \int\dif x \int \dif\kk\;\hz_\pq(\kk)V^{(0)}_\pg(\kk,x)\fz_\pg(x)
\end{equation}
in terms of the LO jet vertex
\begin{equation}\lab{dV0}
 V^{(0)}_\pg(\kk,x) \dug \hz_\pg(\kk)\Sj2(\kk;x)\;,
\end{equation}
which is just the product of the LO gluon impact factor
\begin{equation}\lab{dh0}
 \hz_\pg(\kk) \dug \N\,\frac{C_A}{\kk^2}\;,\quad 
 \N = \frac{2^{1+\e} \as}{\mu^{2\e}\Gamma(1-\e)\sqrt{N_c^2-1}}\;,
\end{equation}
and of the jet distribution for two particles in the final state
\begin{align}
  \Sj2(\kk;x) &\dug S_J^{(2)}(p_1,p_2;p_\pg,p_\pq) \nonumber\\
 &= \delta\left(1-\frac{x_J}{x}\right)E_J^{1+2\e}\delta(\kk-\kk_J)\;,\nonumber\\ \lab{dS2}
 x_J &\dug \frac{E_J\esp{y_J}}{\sqrt s}\;.
\end{align}

\subsection{Ansatz for the factorization formula\labe{s:aff}}

According to the analysis of Ref.~\cite{BaCoVa01}, we propose a high energy factorization
formula for the description of the high energy quark-hadron interaction with a jet in the
final state. In this formula a quark impact factor $h_\pq$, the gluon Green's function $G$,
the gluon-initiated jet vertex $V_\pg$ and the gluon PDF $f_\pg$ are convoluted in both
transverse and longitudinal variables to provide the jet differential cross section as
follows:
\begin{subequations}\labe{ansatz}
\begin{align}\label{FF}
&\dJ{}= \int\dif x \int \dif\kk\,\dif\kk'\;
 h_\pq(\kk) G(xs,\kk,\kk') V_\pg(\kk',x) f_\pg(x) \\ \label{G1loop}
&G(xs,\kk,\kk') \dug \delta(\kk-\kk')+\as K^{(0)}(\kk,\kk')\log\frac{xs}{s_0}
 +\ord(\as^2)\;.
\end{align}
\end{subequations}
The gluon Green's function contains by definition all the energy dependence of the process
and is given in terms of the BFKL kernel~\cite{BFKL76} $K(\kk,\kk')$.
By writing the perturbative expansions for the quark impact factor, the jet vertex,
and the PDF
\begin{subequations}\labe{svil}
\begin{align}\label{svilh}
 h &= \hz + \as h^{(1)}+\cdots \\ \label{svilV}
 V &= V^{(0)} + \as V^{(1)}+\cdots \\ \label{svilf}
 f &= \fz + \as f^{(1)}+\cdots\;,
\end{align}
\end{subequations}
our ansatz corresponds to the following structure for the one-loop cross section:
\begin{align}\nonumber
 \dJ{^{(1)}} &= \as\int\dif x \int \dif\kk\; \bigg\{ 
 \int\dif\kk'\left[\hz_\pq(\kk) K^{(0)}(\kk,\kk')\right.\\ \nonumber
 &\quad\left.\times\log\frac{xs}{s_0} 
 V_\pg^{(0)}(\kk',x) \fz_\pg(x) \right]\\ \nonumber
 &\quad+ h^{(1)}_\pq(\kk)  V_\pg^{(0)}(\kk,x) \fz_\pg(x) \\ \nonumber
 &\quad+ \hz_\pq(\kk)  V_\pg^{(0)}(\kk,x) f^{(1)}_\pg(x)\\ \lab{dJ1loop}
 &\quad+ \hz_\pq(\kk)  V_\pg^{(1)}(\kk,x) \fz_\pg(x) \bigg\}\;,
\end{align}
\noindent
which is obtained simply by expanding Eq.~(\ref{FF}) up to relative order $\as$.

For the first order correction to the partonic impact factor, $h^{(1)}$, which appears on
the second line, we can use the known expression of Ref.~\cite{Ci98,CiCo98}, and for the
correction to the PDF, $f^{(1)}$, we have the usual convolution with the LO Altarelli-Parisi
splitting functions:
\begin{align}\nonumber
 \as f^{(1)}_\pa(x,\mu_F^2) &\dug
 \frac\as{2\pi}\,\frac1\e\left(\frac{\mu_F^2}{\mu^2}\right)^\e \sum_\pb\int_x^1\frac{\dif\xi}{\xi}\;
 P_{\pa\pb}(\xi) \fz_\pb\left(\frac{x}{\xi}\right) \\ \lab{deff1}
 &= \frac\as{2\pi}\,\frac1\e\left(\frac{\mu_F^2}{\mu^2}\right)^\e 
 \sum_\pb P_{\pa\pb}\otimes \fz_\pb\;.
\end{align}
Because of the definition (\ref{defas}) for $\as=g^2\mu^{2\e}/4\pi\,[1+\e(\gamma_E-\log4\pi)]$,
Eq.~(\ref{deff1}) defines the one-loop PDF in the $\MS$ scheme.
Finally, the correction term $V_\pg^{(1)}$ is what we want to compute in this paper.

Equations~(\ref{ansatz}) and (\ref{svil}) constitute a highly non trivial ansatz, which will be
shown to depend upon a careful separation of singular and finite pieces. Our main task will
consist to identify the collinear singularities~(\ref{deff1}) to be absorbed in the parton
densities, to check cancellation of the remaining infrared singularities, and, finally, to
separate the terms proportional to $\log s$ which belong into the first line
of~(\ref{dJ1loop}). The remaining finite (in $\e$) and constant (in $s$) term will
eventually be interpreted as one-loop correction to the jet vertex, $V_\pg^{(1)}$.

\section{Virtual corrections\labe{s:virt}}

For the one-loop analysis of the gluon-initiated jet production
process we adopt dimensional regularization in $D=4+2\epsilon$ dimensions and define,
according to the $\MS$ scheme, the bare dimensionless coupling $\as$ as a function of the
dimensionful bare coupling $g$ and of the renormalization scale $\mu$ as follows:
\begin{equation}\lab{defas}
 \as = \as^\bare \dug \frac{g^2 \mu^{2\e}\Gamma(1-\e)}{(4\pi)^{1+\e}}   
\end{equation}
The one-loop analysis of the virtual corrections can be carried out in the same way of the
quark-initiated case presented in~\cite{BaCoVa01}. Discarding all terms suppressed by powers of
$s$, the one-loop quark-gluon cross section can be derived from Ref.~\cite{PPRcorr} and
the ensuing virtual contribution to the jet cross section reads
\begin{align}\nonumber
 \dJ{^{(\rm virt)}} &= \as\int\dif x\int\dk\;\hz_\pq(\kk)\\ \nonumber
 &\quad\times \left[2\omega^{(1)}(\kk)\log\frac{xs}{\kk^2}
 +\widetilde\Pi_\pq(\kk)+\widetilde\Pi_\pg(\kk)\right]\\ \lab{dJvirt}
 &\quad\times V^{(0)}_\pg(\kk,x) \fz_\pg(x)\;,
\end{align}
$\hz_\pq(\kk) = C_F/C_A \hz_\pg(\kk)$ being the LO quark impact factor.  The first term
represents the leading $\log s$ (LL) contribution to the virtual corrections. The coefficient of
$\log s$, namely $2\omega^{(1)}$, constitutes the virtual part of the leading BFKL kernel
and is just twice the one-loop Regge-gluon trajectory
\begin{equation}\lab{traj}
\omega^{(1)}(\kk) = -\frac{C_A}{\pi}\frac{1}{2\e}\frac{\Gamma^2(1+\e)}{\Gamma(1+2\e)}
\left( \frac{\kk^2}{\mu^2}\right)^\e\;.
\end{equation}
The $\e$-pole reflects the presence of a soft singularity which will be compensated by an
opposite one in the real part of the kernel.

The non logarithmic terms in Eq.~(\ref{dJvirt}) represent the next-to-leading $\log s$ (NLL)
contribution to the virtual corrections after the subtraction of the UV $\e$-pole occurring
in the renormalization of the coupling
\begin{equation}\lab{run}
 \as(\mu^2) \dug \as^\bare\left[1-\as^\bare\frac{b_0}{\e}\right]\;.
\end{equation}
They are expressed in terms of the virtual corrections to the impact factors denoted by
$\widetilde\Pi$.  The virtual correction to the gluon impact factor reads
\begin{align}\nonumber
 \widetilde\Pi_{\pg}(\kk) 
&=\left[\left(-\frac1{\e^2}+\frac{11}{6\e}
 +\frac{5\pi^2}{12}-\frac{67}{36}\right)\frac{C_A}{\pi}\right.\\ \lab{rengIF}
&\quad\left. +\left(-\frac1{3\e}+\frac{5}{18}\right)\frac{\Nf}{\pi}
 -b_0\log\frac{\kk^2}{\mu^2}\right]\left(\frac{\kk^2}{\mu^2}\right)^\e\;,
\end{align}
where $b_0 = (11C_A - 2\Nf)/12\pi$ is the first coefficient of the $\beta$-function.
Any occurrence of $\as$ in Eq.~(\ref{dJvirt}) and in all other coming formulae is to be
understood as $\as(\mu^2)$.

The gluon impact factor virtual correction~(\ref{rengIF}) shows double and
single poles in $\e$. These poles are of IR origin and are due to both soft and collinear
singularities. Partly they will cancel against the corresponding singularities of the real
emission corrections, leaving a simple pole that will be absorbed in the redefinition of the
PDFs. This will be shown in Sec.~\ref{s:jv}.

\section{Real corrections\labe{s:rc}}

The real emission corrections of the quark-gluon collision that we are considering, involves
three partons in the final states: an outgoing quark (labelled by ``2'') --- which is
nothing but the scattered incoming quark $\pq$ --- and two additional partons (``1'' and
``3''), which can be either a $\pq\bar\pq$ pair or two gluons (see Fig.~\ref{f:gq123}).
\begin{figure}[ht!]
\centering
\resizebox{0.2\textwidth}{!}{\includegraphics{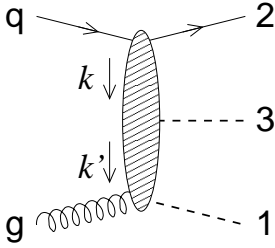}}
\caption{\labe{f:gq123}Labelling the two-to-three parton scattering process. 
  The two dashed lines denote a $\pq\bar\pq$ pair or a gluon pair.
  Note the definition of the transferred momenta $k$ and $k'$.}
\end{figure}

We parametrize the kinematics of the process in terms of Sudakov variables of two exchanged
momenta:
\begin{subequations}\labe{k}
\begin{align}\label{k1}
 k &\dug p_\pq - p_2 = -\bar{w} p_\pg + w p_\pq + k_\perp\;,&
 k_\perp &= (0,0,\kk)\\ \label{k2}
 k' &\dug p_1 - p_\pg = -z p_\pg + \bar{z} p_\pq + k'_\perp\;,&
 k'_\perp &= (0,0,\kk')\;.
\end{align}
\end{subequations}
Note that the transverse energies introduced in Eq.~(\ref{coord3}) correspond to
$E_1=|\kk'|,E_2=|\kk|,E_3=|\kk-\kk'|$.

The partonic differential cross sections has been computed~\cite{Ci98,CiCo98} for both cases
in the high energy regime, where we neglect terms suppressed by powers of $s$.  Introducing
the rapidity $y'=y+\frac12\log\frac1x$ ``measured'' in the partonic center of mass frame, we
can split the phase space into two semispaces defined by $y'_1, y'_3 > 0$
({\em lower half}) and $y'_3<0<y'_1$ ({\em upper half})
\footnote{By momentum conservation there is always a particle with $y'>0$, which can be
  labelled by ``1'' without loss of generality.}.

The form of the partonic differential cross section turns out to be quite simple when
restricted to one of the two halves of the phase space.  For the ``lower half region''
$y'_1,y_3'>0$, which corresponds to $z>z\cut\dug\frac{E_3}{\sqrt{xs}}$, the cross section
can be cast into the general form
\begin{align}\\ \nonumber
 &\dif\hat\sigma_{\pq\pg\to{\it fin}} = \hz_\pq(\kk) F_{\it fin}(\kk,\kk',z) \hz_\pg(\kk')
 \;\dk\, \dk'\, \dif z\;,\\ \lab{realpart}
 &(z>z\cut)\;,
\end{align}
where the function $F$ depends on the particular final state. For quark-gluon scattering, we
have two contributions.  The $\pq\pq\bar{\pq}$ final state term is given by
\begin{align}\lab{Fqqbq}
 F_{\pq\pq \bar{\pq}}(\kk,\kk',z) &= \frac{\as}{2\pi} \Nf T_R 
 \frac{\PP_{\pq\pg}(z,\e)}{\pi_\e}\frac{1}{\qq^2}\\
&\nonumber\quad\times
 \left[ \frac{C_F}{C_A}   + z(1-z)\frac{\qq\cdot \kk'}{(\qq-z \kk)^2}
 \right]\;,\\ \lab{Pqg}
\PP_{\pq\pg}(z,\e) &= 1-\frac{2z(1-z)}{1+\e}\;,\\ \nonumber
 \pi_\e &= \pi^{1+\e}\Gamma(1-\e)\mu^{2\e}\;,
\end{align}
where $\qq=\kk-\kk'$, $T_R=1/2$ and $\PP_{\pg\pq}(z,\e)$ is --- apart
from a missing $T_R$ factor --- the real part of the $\pg\to\pq$ splitting function in
$4+2\e$ dimensions.

The $\pq\pg\pg$ contribution can be written as
\begin{align}\lab{Fqgg}
 F_{\pq\pg\pg}(\kk,\kk',z) &= \frac{\as}{2\pi} C_A
 \frac{\PP_{\pg\pg}(z)}{\pi_\e}\\ \nonumber
&\quad\times
 \frac{z^2 \kk'{}^{2} +(1-z)^2 \qq^2 -z(1-z) \qq \cdot \kk'} {\qq^2 (\qq-z \kk)^2}\;,
 \\ \lab{Pgg}
 \PP_{\pg\pg}(z) &=\frac{1+z^4+(1-z)^4}{2z (1-z)}\;,
\end{align}
where $\PP_{\pg\pg}(z)$ is --- apart from the missing colour factor $2C_A$ --- the real part
of the $\pg \to \pg$ splitting function (in any dimension).

The real correction to the upper quark impact factor receives contribution only from the
$\pq\pg\pg$ final state in the ``upper half region'' of the phase space and from the virtual
correction contribution of Eq.~(\ref{dJvirt}) coming from the $\pq$ impact factor correction
$\widetilde\Pi_\pq$. It can be computed in a similar way as presented in~\cite{BaCoVa01} and
it will not be discussed further here. We simply report the final result adapted to the case
of incoming gluon:
\begin{subequations}\labe{fullnegative}
\begin{align}\nonumber
  &\dJ{^{(y_3'<0)}_{\pq\pg\pg}}+\left.\dJ{^{({\rm virt})}}\right|_{\widetilde\Pi_\pq}\\ \nonumber
  &\quad=\as\int\dif x\int\dk\,\dk'\;\hz_\pq(\kk) K^{(0,{\rm real})}(\kk,\kk')\\
\label{LLnegative}&\qquad\times\log\frac{\sqrt{xs}}{\max(E_2,E_3)}V^{(0)}_\pg(\kk',x) \fz_\pg(x)
  \\ \label{NLLnegative} &\qquad+\as\int\dif x\int\dk\;h^{(1)}_\pq(\kk)
  V^{(0)}_\pg(\kk,x) \fz_\pg(x)\;,
\end{align}
\end{subequations}

The structure of the infrared (IR) singularities for the gluon-initiated vertex, is somewhat
more entangled than that of the quark-initiated one.  This happens in particular when the
gluon pair is emitted in the final state. Since gluons are indistinguishable, each of the
gluons can contribute to the LL term, and the LL subtraction needed to extract the jet
vertex cannot be defined on the same footing as in the quark case~\cite{BaCoVa01}.

In the next sections we develop an explicit procedure for separating the LL contribution
from the collinear singular terms. We check the cancellation of the soft singularities and
isolate the finite expression for the gluon-initiated jet vertex.

\subsection{Jet definition\labe{s:jd}}

We begin with a brief review of the jet definition. Following the arguments given
in~\cite{jetdef}, we require that the jet distribution $S_J^{(n)}$, selecting from a generic
$n$-particle final state the configurations contributing to our one jet inclusive
observable, be IR safe. This corresponds to the fact that emission of a soft particle cannot
be distinguished from the analogous state without soft emission. Furthermore, collinear
emissions of partons cannot be di\-stin\-gui\-shed from the corresponding state where the
collinear partons are replaced by a single parton carrying the sum of their quantum numbers.

Listing the possible IR singular configurations, only the emission of a gluon (1 or 3) with
vanishing momentum gives rise to soft singularities. Collinear singularities arise in
collinear emissions of partons that couple directly to each other. The list of all possible
collinear singular configurations, in an obvious notation, reads as follows:
\begin{subequations}\labe{collconf}
\begin{align}\label{collconf1}
&& &\pg\parallel1\;,&&\pg\parallel3\;,&&1\parallel3\;,&&\\ \label{collconf2}
&& &\pq\parallel1\;,&&\pq\parallel2\;,&&\pq\parallel3\;,&&2\parallel3\;.&&
\end{align}
\end{subequations}
It is important to note that, in the kinematic regime we are considering, configurations in
which quark 2 is emitted outside the fragmentation region of quark $\pq$ are strongly
suppressed, i.e., quark 2 never belongs to the jet produced in the forward direction of
gluon $\pg$. This is because the propagator of the exchanged particle provides a suppression
factor $\sim t/s$ with respect to the situation where quark 2 is in the fragmentation region
of quark $\pq$.  Therefore, we can safely neglect the configurations in which quark 2 enters
the jet, and only particles 1 and 3 play a role in building up the jet.

In terms of the variables $\pp_1, \pp_3, p_3^+, p_\pg^+$, the 3-particle jet distribution
\begin{align}\nonumber
 \Sj3\big(\pp_1,\pp_3,\frac{p_3^+}{p_H^+};\frac{p_\pg^+}{p_H^+}\big) 
 &\equiv \Sj3(\kk',\kk-\kk',xz;x)\\ \lab{defS}
 &\dug S^{(3)}_J(p_1,p_2,p_3;p_\pg,p_\pq)\;. 
\end{align}
has to satisfy the following properties in order to be IR safe (cfr.~\cite{BaCoVa01}):
\begin{subequations}\labe{relS}
\begin{align}\lab{soft1S}
 &\text{1 soft}: & \Sj3(\mbf0,\pp,x;x) &= \Sj2(\pp;x) &&\\ \lab{softS}
 &\text{3 soft}: & \Sj3(\pp,\mbf0,0;x) &= \Sj2(\pp;x) &&\\ \lab{collS}
 &1\parallel3: & \Sj3((1-\lambda)\pp,\lambda\pp,\lambda x;x) &
  = \Sj2(\pp;x) && \\ \lab{factS1}
 &\pg\parallel1: & \Sj3(\mbf0,\pp,\xi;x) &= \Sj2(\pp;\xi) && \\ \lab{factS2}
 &\pg\parallel3: & \Sj3(\pp,\mbf0,\xi;x) &= \Sj2(\pp;x-\xi) \;. &&
\end{align}
\end{subequations}
Note that, since quark 2 does not participate in the jet, the collinear properties of the
jet distribution are applied only to the configurations listed in~(\ref{collconf1}).

In the following sections these relations will be used when extracting the divergences of
the real emission.

\subsection{$\pq\bar\pq$ pair in the final state \labe{s:pss}}

We have already observed in the previous section that there are two kind of final states,
differing in the type of the partons denoted by 1 and 3: either a $\pq\bar\pq$ pair or a
gluon pair. Here we consider the contribution of the final state with $\pq\bar\pq$ pair.  In
this case, the main contribution occurs when the $\pq\bar\pq$ pair emitted with a small
invariant mass in the fragmentation region of the incoming gluon. In practice, we need only
to consider the ``lower half'' phase space $y'_3<0$ contribution. The corresponding Feynman
diagrams are shown in Fig.~\ref{f:GQfrag}.

\begin{figure*}[ht!]
  \centering
  \resizebox{0.75\textwidth}{!}{\includegraphics{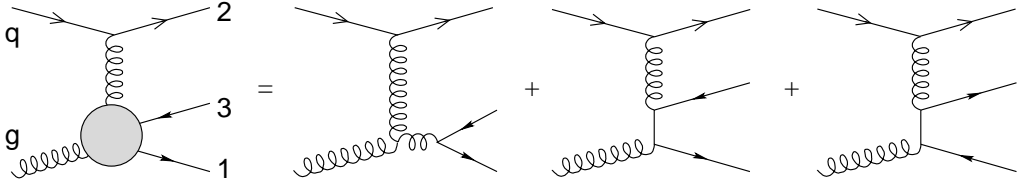}}
  \caption{\labe{f:GQfrag}Feynman diagrams contributing to quark-gluon scattering in which
    the gluon fragments in a $\pq\bar\pq$ pair.}
\end{figure*}

The starting formula is derived from Eq.~(\ref{pSf}), using Eq.~(\ref{defS}) for the jet
distribution and Eqs.~(\ref{realpart}) and~(\ref{Fqqbq}) for the partonic cross section:
\begin{align}\nonumber
 \dJ{_{\pq\pq\bar\pq}} &= \frac{\as}{2\pi}\N \int\dk\,\dk'\;\hz_\pq(\kk)\\ \nonumber
&\quad\times\int_{z\cut}^1\dif z\;\Nf T_R\frac{\PP_{\pq\pg}(z,\e)}{\pi_\e}
 \frac{1}{\kk'{}^2\qq^2} \\ \nonumber
&\quad\times\left[C_F +C_A \frac{z(1-z)\kk'\cdot \qq}{(\qq-z \kk)^2 } \right]\\ \lab{dJqbq}
&\quad\times\int\dif x\; \Sj3\big(\kk',\qq,xz;x\big) \fz_\pg(x)\;.
\end{align}
Since the integrand in the RHS of Eq.~(\ref{dJqbq}) is regular in the $z\to 0$ limit, we can
replace the lower limit of integration in $z$: $z\cut\to 0$. This amounts to a negligible
error in the high energy limit of order $t/s$.
\subsubsection*{$C_F$ term}

Let us first consider the $C_F$ piece.  In this term the two IR singularities of collinear
origin ($3 \parallel \pg \iff \qq=0$ or $1 \parallel \pg \iff \kk'=0$) can be disentangled
employing a simple fract decomposition:
\begin{equation}\lab{sfrac1}
\frac{1}{\qq^2 \kk'{}^2}=\frac{1}{\qq^2+ \kk'{}^2} \left[\frac{1}{\qq^2}+
\frac{1}{\kk'{}^2} \right]\;,
\end{equation}
and we can write
\begin{align}\lab{dJCF13}
 \dJ{_{C_F}} &= \dJ{_{C_F}^{(1)}} + \dJ{_{C_F}^{(3)}}\\ \nonumber
 \dJ{_{C_F}^{(1)}} &= \int\dif\kk\; \hz_\pq(\kk)
 \int\frac{\dif\kk'}{\pi_\e\kk'{}^2} I(\kk';\kk)\\ \nonumber
 \dJ{_{C_F}^{(3)}} &= \int\dif\kk\; \hz_\pq(\kk)
 \int\frac{\dif\kk'}{\pi_\e(\kk-\kk')^2} I(\kk';\kk)
\end{align}
where the integrand $I$ is given by
\begin{align}\nonumber
  I(\kk';\kk)&\dug \N\frac{C_F T_R \Nf}{2\pi}\frac1{\kk'{}^2+(\kk-\kk')^2}\int_0^1
 \dif z\;\PP_{\pq\pg}(z,\e)\\ \lab{defI}
 &\quad\times\int\dif x\;\Sj3\big(\kk',\kk-\kk',xz;x\big) \fz_\pg(x)\;.
\end{align}
Because of the symmetry of $\PP_{\pq\pg}$ and $\Sj3$ under the $\pq\lra\bar\pq$ exchange,
i.e., $\{z\lra1-z,\kk'\lra\kk-\kk'\}$, it holds $I(\kk';\kk)=I(\kk-\kk';\kk)$ and thus the
two terms in Eq.~(\ref{dJCF13}) are equal.

The separation of the collinear singularity is performed by means of the subtraction method
by decomposing
\begin{equation} \lab{dJCF}
 \dJ{_{C_F}} \dug 
 \dJ{^{\rm coll}_{C_F}}+\dJ{^{\rm finite}_{C_F}}\;,
\end{equation}
where the collinear singular term depends on an UV cut-off $\Lambda$ and is defined by
\begin{equation}\lab{CFcoll}
 \dJ{^{\rm coll}_{C_F}}\dug 2 \int\dif\kk\; \hz_\pq(\kk)
 \int\frac{\dif\kk'}{\pi_\e\kk'{}^2} I(\mbf0;\kk)\Theta(\Lambda^2-\kk'{}^2)\;.
\end{equation}

In both collinear limits $1 \parallel \pg$ and $3 \parallel \pg$ the jet distribution
reduces to $\Sj3 \to \Sj2(\kk;x(1-z))$ (cfr.\ Eq.~(\ref{factS1},e)).  Introducing
the gluon-to-quark splitting function $P_{\pq\pg}(z)=T_R [z^2 +(1-z)^2]$ one can write
\begin{align}\nonumber
 \dJ{^{\rm coll}_{C_F}}&= \frac{\as}{\pi}\Nf \int\dk\;\hz_\pq(\kk) \int\dif x
 \\ \nonumber
&\quad\times\Bigl[\frac{1}{\e} \left( \frac{\Lambda^2}{\mu^2}\right)^\e \int_0^1\dif z
 \;P_{\pq\pg}(z) V^{(0)}_{\pq}(\kk,xz)  \fz_\pg(x) \\ \lab{dJCFcoll}
&\quad+ \int_0^1\dif z\; z(1-z) V^{(0)}_{\pq}(\kk,xz)  \fz_\pg(x)\Bigr] .
\end{align}
The finite contribution of the $C_F$ term can be computed at $\e=0$ and reads
\begin{align}\nonumber
\dJ{^{\rm finite}_{C_F}}&= \frac{\as}{\pi}\Nf\int\dk\;\hz_\pq(\kk)\int \frac{\dk'}{\pi}
\int\dif x \int_0^1\dif z\;P_{\pq\pg}(z)\\ \nonumber
&\quad\times\Bigl[ \hz_\pq(\kk') \frac{1}{\qq^2 + \kk' {}^2} \Sj3\big(\kk',\qq,xz;x\big)
\\ \lab{dJCFfinite}
&\quad-\frac{1}{\kk'{}^2} \Theta(\Lambda^2 -\kk'{}^2) V^{(0)}_{\pq}(\kk,xz) \Bigr]  \fz_\pg(x).
\end{align}

\subsubsection*{$C_A$ term}

In the $C_A$ term there is only one final state collinear singularity when $3 \parallel 1$,
corresponding to $\qq-z \kk=0$.  Also in this case we separate the collinear singularity
\begin{equation} \lab{dJCA}
\dJ{_{C_A}} \dug \dJ{^{\rm coll}_{C_A}}+\dJ{^{\rm finite}_{C_A}}
\end{equation}
by means of the subtraction method: we define the col\-li\-near singular term as the residue
of the $1/(\qq-z \kk)^2$ pole, integrated in a circle of radius $\Lambda$ centered in the
singularity. For simplicity we use the same cut-off $\Lambda$ introduced in the $C_F$ term.
In this case, the jet distribution can be simplified thanks to Eq.~(\ref{collS}): $\Sj3 \to
\Sj2(\kk;x)$ and the $z$ integration can straightforwardly be performed, yielding
\begin{align}\nonumber
\dJ{^{\rm coll}_{C_A}}&= \as\int\dk\;\hz_\pq(\kk)\int\dif x\;
 V^{(0)}_\pg(\kk,x) \fz_\pg(x)\\ \lab{dJCAcoll}
&\quad\times \frac{\Nf}{\pi}\left( \frac{\Lambda^2}{\mu^2}\right)^\e
 \left( \frac{1}{6\e} + \frac{1}{12}\right)
\end{align}
and
\begin{align}\nonumber
\dJ{^{\rm finite}_{C_A}}&= \frac{\as}{2\pi} \Nf 
\int\dk\;\hz_\pq(\kk)\int\dif x\;\fz_\pg(x)\\ \nonumber
&\quad\times\int_0^1\dif z\;P_{\pq\pg}(z) \int\frac{\dk'}{\pi} \frac{\N C_A}{(\qq-z \kk)^2}
\\ \nonumber
&\quad\times\Bigl[ z(1-z) \frac{\qq \cdot \kk'}{\qq^2 \kk' {}^2}  \Sj3\big(\kk',\qq,xz;x\big)
\\ \lab{dJCAfinite}
&\quad-\frac{1}{\kk^2}\Theta\big(\Lambda^2-(\qq-z \kk)^2\big) \Sj2(\kk;x)\Bigr]
\end{align}

\subsection{Real corrections to the jet vertex from $\pq\pg\pg$ final state \labe{s:pgr}}

The starting formula can be derived from Eq.~(\ref{pSf}), using Eq.~(\ref{defS}) for the jet
distribution and Eqs.~(\ref{realpart}) and~(\ref{Fqgg}) for the partonic cross section. A
more convenient expression can be written if one notes that the splitting function for
gluons can be decomposed as
\begin{equation}
  \PP_{\pg\pg}(z) =P(z)+P(1-z) , \quad 
 P(z)=\left(\frac{1}{z}+\frac{z}{2}\right) (1-z).
\end{equation} 
Using the symmetry of the partonic cross section and of the jet distribution under the
exchange of the two gluons: $\{z\lra1-z,\kk'\lra\kk-\kk'\}$, the jet cross section can be
rewritten as
\begin{align}\nonumber
 \dJ{^{(y_3'>0)}_{\pq\pg\pg}} &= \frac{\as}{\pi}\int\dk\; \hz_\pq(\kk') \int \frac{\dk'}{\pi_\e}\;
 \hz_\pg(\kk) \int_{z\cut}^1\dif z\; P(z)\\ \nonumber
&\quad\times\frac{z^2 \kk'{}^2 +(1-z)\qq \cdot (\qq-z \kk)}{\qq^2 (\qq-z \kk)^2 }\\ \lab{dJgg}
&\quad\times\int\dif x\; \Sj3\big(\kk',\qq,xz;x\big) \fz_\pg(x)\;.
\end{align}
It is important to observe that now the ``splitting function'' $P$ has a pole only at
$z=0$, while it is regular (actually vanishes) at $z=1$. The reason is that now we are
employing an asymmetric treatment which causes gluon 3 to be the only responsible of soft
singularities and of the central region LL contribution.

The expression (\ref{dJgg}) looks now pretty similar to the real contribution of the
quark-initiated case.  We can therefore proceed in an analogous way for its computation and
consider separately the two terms in (\ref{dJgg}). We define the $A$ and $B$ term as the
first ($\propto z^2 \kk'{}^2$) and the second ($\propto (1-z)\qq \cdot (\qq-z \kk)$) terms
of Eq.~(\ref{dJgg}) respectively.

\subsubsection{$A$ term}

The $A$ term%
\footnote{This term has the same structure of the $C_F$ term analyzed in
  Ref.~\cite{BaCoVa01}, and can be simply recovered from the latter by replacing
  $C_F^2\PP_{\pg\pq}\to C_A^2 P$. The analysis that we perform here closely follows the one
  made in that paper.}%
, owing to the $\kk'{}^2$ factor in the numerator, has no $\pg\parallel 1$ collinear
singularity. In addition, the $z^2$ factor in the numerator causes a suppression of the
integrand in the central region, and we can shift the lower limit of integration $z\cut$ in
$z$ down to zero:
\begin{align} \nonumber
 \dJ{_{A}} &= \as\frac{C_A}{\pi}\int\dk\;\hz_\pq(\kk) \int\dif x\;\fz_\pg(x)
 \int_0^1\dif z\;z^2 P(z) \\\lab{A1}
&\quad\times
 \int\frac{\dk'}{\pi_\e}\frac{\N C_A}{\qq^2(\qq-z\kk)^2}\Sj{3}(\kk',\qq,xz;x)\;.
\end{align}
We perform the rescaling $\qq\ugd z\lb$ and use $\lb$ as integration variable by
substituting $\kk'=\kk-z\lb$, so that $\qq-z\kk=z(\lb-\kk)$. Next we perform a simple
fraction decomposition in order to separate the initial ($i$) state ($\pg \parallel
3\iff\lb=0$) and final ($f$) state ($1 \parallel 3\iff\lb-\kk=0$) collinear singularities:
\begin{equation}\lab{fract}
 \frac1{\lb^2(\lb-\kk)^2} = \frac1{\lb^2+(\lb-\kk)^2}\left[\frac1{\lb^2}+\frac1{(\lb-\kk)^2}
 \right]\;.
\end{equation}

Beginning with the {\bf final state} ($f$) collinear singularity, in terms of the new
variables the $A$ contribution to the jet cross section can be rewritten in the form
\begin{align}\nonumber
 \dJ{^{f}_{A}} &\dug \as\int_0^1\frac{\dif z}{z^{1-2\e}}
 \int\frac{\dif\lb}{\pi_\e(\lb-\kk)^2} I(z,\lb)\\ \lab{dJAf}
&= \dJ{^{f,{\rm soft}}_{A}}+\dJ{^{f,{\rm coll}}_{A}}+\dJ{^{f,{\rm finite}}_{A}}\;,
\end{align}
which is particularly suitable for the analytic extraction of the divergences: the RHS
contains three pieces, {\it 1)} the soft divergence, {\it 2)} the collinear divergence, and
{\it 3)} a finite part.  The integrand introduced in Eq.~(\ref{dJAf}) is defined by
\begin{align}\nonumber
 I(z,\lb) &\dug \frac{C_A}{\pi} z P(z)\int\dk\;\hz_\pq(\kk)
 \frac{\N C_A}{\lb^2+(\lb-\kk)^2}\\ \lab{idJAf}
&\quad\times\int \dif x \, \Sj{3}(\kk-z\lb,z\lb,xz;x) \fz_\pg(x)\;.
\end{align}
The soft term in Eq.~(\ref{dJAf}) is defined by evaluating the integrand in the soft limit
$z\to0$. In this limit, the jet distribution can be simplified by means of
Eq.~(\ref{softS}), and leads to the constraint $\kk^2=E_J^2$. One obtains
\begin{align}\nonumber
\dJ{^{f,{\rm soft}}_A} &\dug \as\int_0^1\frac{\dif z}{z^{1-2\e}}
 \int\frac{\dif\lb}{\pi_\e(\lb-\kk)^2} I(0,\lb) \\ \nonumber
&= \as\frac{C_A}{\pi}\int\dk\;\hz_\pq(\kk) \int_0^1 \frac{\dif z}{z^{1-2\e}}
\int \frac{\dif \lb}{\pi_\e(\lb-\kk)^2}\\ \nonumber
&\quad\times\frac{\N C_A}{\lb^2+(\lb-\kk)^2}
\int \dif x \; \Sj{2}(\kk; x) \fz_\pg(x) \\ \nonumber
&= \as\frac{C_A}{\pi} \left[\frac{1}{2\e^2} -\frac{\pi^2}{12}  \right]
 \left(\frac{E_J^2}{\mu^2} \right)^\e\\ \lab{Afsoft}
&\quad\times\int\dk\int\dif x\;\hz_\pq(\kk) V^{(0)}_\pg(\kk,x)\fz_\pg(x)\;,
\end{align}
where as usual, in the final result,
we have collected some factors in such a way that they reproduce
the LO jet vertex~(\ref{dV0}). The divergent factor exhibits single as well as
double poles, because our definition of the soft part includes also the region where
collinear and soft singularities merge.

The pure collinear singularity can be isolated by evaluating the integrand~(\ref{idJAf}) in
the collinear limit $\lb=\kk$, after having subtracted the soft term $(\lb=\kk,z=0)$.  The
resulting expression is clearly regular in the soft limit ($z \to 0$) and therefore contains
a simple collinear pole.  An UV cutoff $\Lambda$ is introduced since the residue at the
collinear limit is no more integrable in the UV region.  
Thanks to Eq.~(\ref{collS}) the jet
distribution simplifies to $\Sj2$:
\begin{align}\nonumber
 \dJ{^{f,{\rm coll}}_A} &\dug \as\int_0^1 \frac{\dif z}{z^{1-2\e}}
 \int \frac{\dif \lb}{\pi_\e(\lb-\kk)^2}\\ \nonumber
&\quad\times\left[ I(z,\kk)-I(0,\kk) \right]
 \Theta(\Lambda^2-(\lb-\kk)^2) \\ \nonumber
&= \as\frac{C_A}{\pi}\int\dk\;\hz_\pq(\kk) \frac{\N C_A}{\kk^2}
 \int_0^1 \frac{\dif z}{z^{1-2\e}}\\ \nonumber
&\quad\times\left[ z P(z) -1 \right]\int\frac{\dif \lb}{\pi_\e(\lb-\kk)^2}
 \Theta(\Lambda^2-(\lb-\kk)^2)\\ \nonumber
&\quad\times\int \dif x \; \Sj{2}(\kk; x) \fz_\pg(x) \\ \nonumber
&= \as\frac{C_A}{\pi}\left[-\frac{11}{12 \e}\left(\frac{\Lambda^2}{\mu^2}\right)^\e+
\frac{67}{36} \right]\\  \lab{Afcoll}
&\quad\times\int\dif x\int \dk\;\hz_\pq(\kk) V^{(0)}_\pg(\kk,x)\fz_\pg(x) \;.
\end{align}
The remaining part is regular in the $\e\to0$ limit and defines the finite term:
\begin{align}\nonumber
 \dJ{^{f,{\rm finite}}_A} \hspace{-3pt}&\dug \as\int_0^1 \frac{\dif z}{z}
 \int\frac{\dif \lb}{\pi(\lb-\kk)^2}\big[ I(z,\lb)-I(0,\lb)\\ \nonumber
&\quad -\big( I(z,\kk)-I(0,\kk)\big)\Theta(\Lambda^2-(\lb-\kk)^2) \big]\\ \nonumber
&=\as\frac{C_A}{\pi} \int \dk\;\hz_\pq(\kk)
\int_0^1 \frac{\dif z}{(1-z)_+}\\ \nonumber
&\quad\times \left[(1-z) P(1-z)\right]
 \int\dif x\;\fz_\pg(x)\\ \nonumber
&\quad\times\int\frac{\dif \lb}{\pi\lb^2}\Bigl[\frac{\N C_A}{\lb^2+(\lb -\kk)^2}\\ \nonumber
&\quad\times\Sj{3}(\lb+z(\kk-\lb),(1-z)(\kk-\lb),x(1-z);x)\\ \lab{Affinite}
&\quad-V^{(0)}_\pg(\kk,x)\Theta(\Lambda^2-\lb^2) \Bigr]
\end{align}
In the last equation we have performed the change of variable $z\to1-z$ in order to simplify
the expression with the introduction of the $()_+$ regularization for regularizing the
$1/(1-z)$ distribution at $z=1$.

Next we consider the term ($i$) with the {\bf initial state} collinear singularity. We can
write, in the same way as before,
\begin{align}\nonumber
 \dJ{^{i}_A} &\dug
 \as\int_0^1 \frac{\dif z}{z^{1-2\e}} \int \frac{\dif \lb}{\pi_\e \lb^2} I(z,\lb)\\ \lab{dJAi}
 &= \dJ{^{i,{\rm soft}}_A}+\dJ{^{i,{\rm coll}}_A}+\dJ{^{i,{\rm finite}}_A} \;,
\end{align}
where $I$ is given by Eq.~(\ref{idJAf}).
It is trivial to see that the soft contribution, defined by evaluating the integrand $I$ in
Eq.~(\ref{dJAi}) at $z=0$, is equal to the corresponding ($f$) term~(\ref{Afsoft}):
\begin{equation} \lab{Aisoft}
 \dJ{^{i,{\rm soft}}_A} = \dJ{^{f,{\rm soft}}_A}\; .
\end{equation}
As to the collinear piece, we note that in the collinear limit $\lb=0$ the jet distribution
reduces (by applying Eq.~(\ref{factS2})) to $\Sj2$, and one gets, using always the same UV
cutoff $\Lambda$,  performing the $\lb$ integration
and changing the integration variable $z \to 1-z$,
\begin{align}\nonumber
 \dJ{^{i,{\rm coll}}_A} &\dug \as\int_0^1 \frac{\dif z}{z^{1-2\e}}
 \int \frac{\dif \lb}{\pi_\e \lb^2}\\ \nonumber
&\quad\times\left[ I(z,\mbf{0})-I(0,\mbf{0}) \right]
 \Theta(\Lambda^2-\lb^2) \\ \nonumber
&= \as\frac{C_A}{\pi}\int \dk\;\hz_\pq(\kk) \int \dif x \; \fz_\pg(x)\\ \nonumber
&\quad\times\Biggl\{ \frac{1}{\e} \left( \frac{\Lambda^2}{\mu^2} \right)^\e 
\int _0^1 \frac{\dif z}{(1-z)_+} (1-z)P(1-z)\\ \nonumber
&\quad\times V^{(0)}_\pg(\kk,x z) + 2 \int _0^1\dif z \left[\frac{\ln(1-z)}{(1-z)}\right]_+
 \nonumber \\ \lab{Aicoll}
&\quad\times(1-z)P(1-z) V^{(0)}_\pg(\kk,x z) 
\Biggr\}.
\end{align}

The left contribution is regular in 4 dimensions and defines another finite term
\begin{align}\nonumber
\dJ{^{i,{\rm finite}}_A} &\dug \as\int_0^1 \frac{\dif z}{z}
 \int\frac{\dif \lb}{\pi \lb^2}\big[ I(z,\lb)-I(0,\lb)\\ \nonumber
&\quad -\big( I(z,0)-I(0,0)\big)\Theta(\Lambda^2-\lb^2) \big] \\ \nonumber
&=\as\frac{C_A}{\pi} \int \dk\;\hz_\pq(\kk)
\int_0^1 \frac{\dif z}{(1-z)_+} \\ \nonumber
&\quad\times\left[(1-z) P(1-z)\right]
 \int\dif x\;\fz_\pg(x)\\ \nonumber
&\quad\times \int \frac{\dif \lb}{\pi \lb^2}\;\Bigl[ \frac{\N C_A}{\lb^2+(\lb -\kk)^2}
\\ \nonumber
&\quad\times\Sj{3}(\kk-(1-z)\lb,(1-z)\lb,x(1-z);x)\\ \lab{Aifinite}
&\quad- V^{(0)}_\pg(\kk,x z) \Theta(\Lambda^2-\lb^2) \Bigr].
\end{align}

\subsubsection{$B$ term}

The last piece to be analyzed is the $B$ term, which reads
\begin{align}\nonumber
 \dJ{_B} &= \as\frac{C_A}{\pi}\int\dif\kk\;\hz_\pq(\kk)\int\dif x\;\fz_\pg(x)
 \\ \nonumber
 &\quad\times\int_{z\cut}^1\dif z\;(1-z) P(z) \int\frac{\dif\kk'}{\pi_\e \kk'{}^2}\\ \nonumber
&\quad\times\frac{\qq\cdot(\qq-z\kk)}{\qq^2(\qq-z\kk)^2}\Sj3(\kk',\qq,xz;x) \\ \lab{B1}
 &= \dJ{_B^{\rm coll}} + \dJ{_B^{\rm LL}} + \dJ{_B^{\rm const}}\;,
\end{align}
and will be decomposed in a collinear divergent piece, an energy dependent piece which is
just the LL contribution to the cross section coming from gluon emission in the central
region, plus a finite and constant in energy term as shown in the last line of the above
equation.  It can be noticed immediately the presence of a $\pg\parallel1$ collinear
singularity corresponding to the $\kk'=0$ pole. As already commented for the similar
case~\cite{BaCoVa01} there are no other singularities in the $\kk'$-integration (neither
$\qq=0$ nor $\qq-z\kk=0$) except for $z\to0$ corresponding to gluon 3 being in the central
region, which gives the LL contribution. Here we expect a soft singularity needed to cancel the
one in the gluon trajectory~(\ref{traj}).

We recall that the jet distribution functions become essential in disentangling the collinear
singularities, the soft singularities, and the leading $\log s$ pieces.

The basic mechanism are the same as for the quark case (remember here that gluon 3 is
playing a special role because of the manipulation in the integrand performed at the
beginning of Sec.~\ref{s:pgr}):
\begin{itemize}
\item When the outgoing gluon 1 is in the collinear region of the incoming gluon $\pg$,
  i.e., $y_1\to\infty$, it cannot enter the jet; only gluon 3 can thus be the jet, $y_3$ is
  fixed and no logarithm of the energy can arise due to the lack of evolution in the gluon
  rapidity. No other singular configuration is found when $J=\{3\}$.
\item In the composite jet configuration, i.e., $J=\{1,3\}$, the gluon rapidity is bounded
  within a small range of values, and also in this case no $\log s$ can arise. There could
  be a singularity for vanishing gluon 3 momentum: even if the $1\parallel3$ collinear
  singularity is absent, we have seen that, at very low $z$, a soft singular integrand
  arises. However, the divergence is prevented by the jet cone boundary, which causes a
  shrinkage of the domain of integration $\sim z^2$ for $z\to0$ and thus compensates the
  growth of the integrand.
\item The jet configuration $J=\{1\}$ allows gluon $3$ to span the whole phase space, apart,
  of course, from the jet region itself. The LL term arises from gluon
  con\-fi\-gu\-ra\-tions in the central region. Therefore, it is crucial to understand to
  what extent the differential cross section provides a leading contribution. It turns out
  that the coherence of QCD radiation suppresses the emission probability for gluon 3
  rapidity $y_3$ being larger than the rapidity $y_1$ of the gluon 1, namely an angular
  ordering prescription holds. This will provide the final form of the leading term, i.e.,
  the appropriate scale of the energy and, as a consequence, a finite and definite
  expression for the one-loop jet vertex correction.
\end{itemize}

Let us isolate in Eq.~(\ref{B1}) the initial state $\pg\parallel1$ collinear singular
contribution and define, as usual, the collinear term by setting $\kk'=0$ (except in the
$1/\kk'{}^2$ pole), and by introducing an UV cutoff.  Observing that the jet distribution,
because of Eq.~(\ref{factS1}), reduces to $\Sj2(\kk;xz)$ one easily obtains
\begin{align}\nonumber
 \dJ{_B^{\rm coll}} &= \as\frac{C_A}{\pi}\int\dif x\int\dk\;\hz_\pq(\kk)\int_0^1\dif z\;
 V^{(0)}_\pg(\kk,xz)\\ \lab{Bcoll}
&\quad\times\fz_\pg(x) \left[\frac{1}{\e}\left(\frac{\Lambda^2}{\mu^2}\right)^\e 
 P(z) \right]\;.
\end{align}

The LL part can be extracted exactly with the same procedure already followed in
\cite{BaCoVa01}, thanks to the suppression of the partonic cross section when gluon 3 is
emitted at larger rapidity w.r.t. that of gluon 1.  Infact, when gluon 3 is in the central
region, gluon 1 must be the jet. In this case, the azimuthal average of the cross section
w.r.t. gluon 3 azimuthal angle $\phi_3$ at fixed gluon 1 momentum yields
\begin{equation}\label{average}
 \left\langle\frac{(1-z)\qq\cdot(\qq-z\kk)}{\qq^2(\qq-z\kk)^2}\right\rangle_{\phi_3}
 = \frac1{\qq^2}\Theta(E_3-z(E_1+E_3))\;.
\end{equation}
This relation is exact and clearly shows that, outside the angular ordered region
\begin{equation}\label{angordpos}
 \frac{E_3}{z} > \frac {E_1}{1-z} \quad\iff\quad \theta_3 > \theta_1 
 \quad\iff\quad y_3 < y_1\;,
\end{equation}
there is no contribution to the cross section. In practice, by taking into account the
variation of $\hz_\pq(\kk)$ during the averaging procedure, instead of a strictly vanishing
contribution we have a strong suppression. Note that, in the limit $\qq\to0$ (which
includes the soft region), the variation of $\kk'$ goes to zero as well, so that
Eq.~(\ref{angordpos}) is really an accurate statement in the ``dangerous'' part of the phase
space. Moreover, Eq.~(\ref{average}) shows that the $1/\qq^2$ kinematic dependence of the LL
kernel governs the differential cross section up to the very end of the angular boundary.

Therefore, we define the LL contribution in the ``lower half region'' $y_3'>0$ by
\begin{align}\nonumber
 \dJ{_B^{\rm LL}} &\dug \as\int\dk\;\hz_\pq(\kk)\int\dk'\;
 \frac{C_A}{\pi}\frac1{\pi_\e \qq^2}\hz_\pg(\kk') \\ \nonumber
&\quad\int_{z\cut}^1\frac{\dif z}{z}\;
 \Theta(E_3-z(E_1+E_3))\\ \nonumber
&\quad\times\int\dif x\;\Sj{2}(\kk', x)\fz_\pg(x) \\ \nonumber
&= \int\dif x\int\dk\int\dif\kk'\;\hz_\pq(\kk) K^{(0,{\rm real})}(\kk,\kk')\\ \lab{Bkern}
&\quad\times\log{\frac{\sqrt{xs}}{E_J+E_3}} V^{(0)}_\pg(\kk',x) \fz_\pg(x)\;,
\end{align}
having imposed the jet condition $J=\{1\}$.

The remaining part of the $B$ term is finite in 4 dimensions and constant in energy, so
that we can set $\e=0$ and $z\cut=0$ to define the constant part
\begin{equation} \lab{Bconst}
 \dJ{_B^{\rm const}} \dug \left[\dJ{_B} - \dJ{_B^{\rm coll}} - \dJ{_B^{\rm LL}}
 \right]{\!\!}_{\mscr{\begin{array}{r} z\cut=0  \\ \e=0 \end{array}}}\;.
\end{equation}

\section{The NLO jet vertex: sum of real and virtual corrections\labe{s:jv}}

Having completed the calculation of both the virtual and real corrections in the whole phase
space, we are going to collect all partial results and to show that the complete one-loop
jet cross section can naturally be fitted to the form of Eq.~(\ref{dJ1loop}).
Table~\ref{t:albero} summarizes the decomposition of the one-loop jet cross section and
gives the references of the various contributions.

In Sec.~\ref{s:virt} we have presented the virtual contributions to the jet cross section
which, after renormalization of the coupling, assume the form of Eq.~(\ref{dJvirt}).
The contribution coming from the $\widetilde\Pi_\pq$ impact
factor correction has been combined with the ``upper half region'' real
contribution to give the full impact factor of the upper quark $\pq$.

The remaining virtual terms  and the LL contributions of both the negative
and positive rapidity regions (the latter contribution given in (\ref{Bkern}) )
can be conveniently rewritten in the form
\begin{align}\nonumber
&\left.\dJ{^{({\rm virt})}}\right|_{\omega^{(1)}}+\left.\dJ{^{({\rm virt})}}
 \right|_{\widetilde\Pi_\pq}+\left.\dJ{^{({\rm real})}}
 \right|_{\rm LL}=\\ \nonumber
&=\as\int\dif x\int\dif\kk\,\dif\kk'\;\hz_\pq(\kk) K^{(0)}(\kk,\kk')\\ \lab{LL}
&\quad\times\log\frac{xs}{s_0(\kk,\kk')} V^{(0)}_\pg(\kk',x)\fz_\pg(x) \\ \nonumber
&\quad+\as\left[ \left(\frac{\kk^2}{\mu^2}\right)^\e
\left( -\frac{C_A}{\pi}\frac{1}{\e^2}
 +\frac{11C_A - 2\Nf}{6 \pi}\frac{1}{\e} \right)\right.\\ \nonumber
&\quad\left.+\frac{C_A}{\pi} \left(\frac{5}{12}\pi^2 -\frac{67}{36}\right)+\frac{5}{18}
 \frac{\Nf}{\pi} -b_0 \log \frac{\kk^2}{\mu^2} \right]\\ \lab{virtQ}
&\quad\times\int\dif x\int\dif\kk\;\hz_\pq(\kk) V^{(0)}_\pg(\kk,x)\fz_\pg(x)\;,
\end{align}
where the kernel $K^{(0)}$ is the full BFKL kernel and the energy scale $s_0$ in
Eq.~(\ref{LL}) has been chosen as
\begin{equation}\lab{energyscale}
 s_0(\kk,\kk') \dug (|\kk'|+|\qq|)(|\kk|+|\qq|) = (E_J+E_3)(E_2+E_3)\; .
\end{equation}
\begin{table*}[tp!]
 \caption{\labe{t:albero}Schematics of the decomposition of real and virtual one-loop
   corrections to $\pg\pq$ scattering and references of the corresponding equations.}
 \centering
\begin{tabular}{|cccc|ccc|ccc|ccc|cc|cc|}
\hline
   \multicolumn{3}{|c|}{virtual} &
   \multicolumn{10}{c|}{real $\pq\pg\pg$} &
   \multicolumn{4}{c|}{real $\pq\pq\bar\pq$} \\
\hline \slarga{2.7}{8}
   $ \omega^{(1)}$ &
   $\widetilde\Pi_\pq$ &
   $\widetilde\Pi_\pq$ &
   \multicolumn{1}{|c}{$y_3'<0$} &
   \multicolumn{9}{|c|}{$y_3'>0$} &
   \multicolumn{4}{c|}{} \\
\cline{5-17}
   \multicolumn{2}{|c}{\slarga{2}{7}} &
   \multicolumn{2}{c|}{$\underbrace{\qquad\quad\qquad}{}$} &
   \multicolumn{6}{|c|}{$A$} &
   \multicolumn{3}{c|}{$B$} &
   \multicolumn{2}{c|}{$C_F$} &
   \multicolumn{2}{c|}{$C_A$}\\
\cline{5-17}
   \multicolumn{2}{|c}{\slarga{2}{7}} &
   \multicolumn{1}{r}{$\swarrow$} &
   \multicolumn{1}{l}{$\searrow$} &
   \multicolumn{3}{|c|}{$f$} &
   \multicolumn{3}{|c|}{$i$} &
   \multicolumn{3}{|c|}{} &
   \multicolumn{2}{|c|}{} &
   \multicolumn{2}{|c|}{}\\
\cline{5-10}
  \pic LL &
  \parbox[b]{15pt}{\pic soft coll} &
  \pic $h^{(1)}_\pq$ &
  \pic LL &
  \slarga{2}{7} \pic soft &
  \pic coll &
  \pic fin &
  \pic soft &
  \pic coll &
  \pic fin &
  \pic coll &
  \pic LL &
  \pic const &
  \pic coll &
  \pic fin &
  \pic coll &
  \pic fin
\\
  \pic (\ref{LL}) &
  \pic (\ref{virtQ}) &
  \pic (\ref{NLLnegative}) &
  \multicolumn{1}{c}{\pic (\ref{LLnegative})} &
  \pic (\ref{Afsoft}) &
  \pic (\ref{Afcoll}) &
  \multicolumn{1}{c}{\pic (\ref{Affinite})} &
  \pic (\ref{Aisoft}) &
  \pic (\ref{Aicoll}) &
  \multicolumn{1}{c}{\pic (\ref{Aifinite})} &
  \pic (\ref{Bcoll}) &
  \pic (\ref{LL}) &
  \multicolumn{1}{c}{\pic (\ref{Bconst})} &
  \pic (\ref{dJCFcoll}) &
  \multicolumn{1}{c}{\pic (\ref{dJCFfinite})} &
  \pic (\ref{dJCAcoll}) &
  \pic (\ref{dJCAfinite}) \\
\hline
\end{tabular}
\end{table*}

Let us stress that the energy scale in Eq.~(\ref{energyscale}) arises naturally when one
requires impact factors and PDFs to have standard collinear properties and the remaining
non-leading-log term to be finite in both the physical $\e\to0$ and high-energy $s\to\infty$
limits.  Choosing a different scale of the energy requires the introduction of additional
NLL operators, which has to be added as multiplicative corrections to the Green's function.
If, for instance, we adopt $s_0=|\kk||\kk'|$, then the Green's function (\ref{G1loop})
has to be replaced by
\begin{align}\nonumber
 G(xs,\kk,\kk')&=(\id+\as H_L)\left[\id+\as K^{(0)}\log\frac{xs}{|\kk||\kk'|}\right]
 \\ \lab{HGH} &\quad\times(\id+\as H_R)\\ \nonumber
 H_L(\kk,\kk')&=-K^{(0)}(\kk,\kk')\log\frac{|\kk|+|\qq|}{|\kk|}\\ \lab{defHLHR}
 H_R(\kk,\kk')&=H_L(\kk',\kk)\;.
\end{align}

We now consider the sum of all $\e$-divergent contributions coming from both real and
virtual corrections.  All double poles coming from (\ref{virtQ}), (\ref{Afsoft}) and
(\ref{Aisoft}) cancel out.  The single poles come from (\ref{virtQ}), (\ref{dJCFcoll}),
(\ref{dJCAcoll}), (\ref{Afcoll}), (\ref{Aicoll}), (\ref{Bcoll}).  They combine to give
exactly the DGLAP splitting functions terms which define the PDF one-loop corrections.  One
has only to note that
\begin{equation}
\frac{(1-z)P(1-z)}{(1-z)_+}+P(z) = \frac{z}{(1-z)_+}+\frac{1-z}{z}+z(1-z)
\end{equation}
and that the $P_{\pg\pg}$ LO DGLAP splitting function is defined by
\begin{align}\nonumber
 P_{\pg\pg}(z)&= 2C_A\left[\frac{1-z}{z}+\frac{z}{(1-z)_+}+z(1-z)\right]\\ \lab{Pggfull}
&\quad+\left(\frac{11C_A}{6} -\frac{\Nf}{3}\right)\delta(1-z)\;.
\end{align} 
The non-LL total singular contribution adds up to
\begin{align}\nonumber
 \dJ{^{\rm singular}} &=
 \frac{\as}{2\pi}\frac1{\e}\left(\frac{\Lambda^2}{\mu^2}\right)^\e
 \int\dif x\int \dif\kk\;\hz_\pq(\kk)\\ \nonumber
&\quad\times\left\{ V_\pq^{(0)}(\kk,x)\,\big[2\Nf P_{\pq\pg} \otimes \fz_\pg\big](x)\right.
 \\ \lab{singular}
  &\quad\left.+V_\pg^{(0)}(\kk,x)\,\big[P_{\pg\pg} \otimes \fz_\pg\big](x)\right\}\; .
\end{align}
The cutoff $\Lambda$ introduced in all the collinear subtraction has the physical
interpretation of factorization scale and can therefore be identified with $\mu_F$.
Eq.~(\ref{singular}) shows that the only singular terms remaining after the sum of 
real and virtual corrections are of collinear origin, and actually amount to the
expected collinear singularities stemming from parton radiation out of incoming gluons.
They can be absorbed in the PDFs according to Eq.~(\ref{deff1}); of course only the term
$\pb=\pg$ in the sum is reproduced. 

Finally all the finite and not LL terms from (\ref{virtQ}), (\ref{dJCFcoll}),
(\ref{dJCFfinite}), (\ref{dJCAcoll}), (\ref{dJCAcoll}), (\ref{dJCAfinite}), (\ref{Afsoft}),
(\ref{Aisoft}), (\ref{Afcoll}), (\ref{Aicoll}), (\ref{Affinite}), (\ref{Aifinite}),
(\ref{Bcoll}) and (\ref{Bconst}) contribute to the jet vertex
\begin{equation}\lab{dJV1}
 \dJ{^{\rm finite}} = \as\int\dif x\int\dif\kk\;\hz_\pq(\kk)V_\pg^{(1)}(\kk,x)\fz_\pg(x)\;, 
\end{equation}
which defines the NLO correction to the gluon-initiated jet vertex
\begin{align}\lab{V1}
&V^{(1)}_\pg(\kk,x)\nonumber \\
&\dug
\left[\left( \frac{11}{6}\frac{C_A}{\pi} -\frac{1}{3}\frac{\Nf}{\pi} \right)
\log\frac{\kk^2}{\Lambda^2}+\frac{\pi^2}{4}\frac{C_A}{\pi}+
\frac{13}{36}\frac{\Nf}{\pi}\right. \nonumber\\
&\quad \quad \left.-b_0\log\frac{\kk^2}{\mu^2}\right]
V_\pg^{(0)}(\kk,x)\; \nonumber \\ 
&\quad + \int\dif z\; V_\pg^{(0)}(\kk,x z)
\Bigl[\frac{\Nf}{\pi}\frac{C_F}{C_A} z(1-z)\nonumber\\
&\quad \quad+2 \frac{C_A}{\pi} (1-z)P(1-z) \left(\frac{\log(1-z)}{1-z}\right)_+ \; \Bigr] 
\nonumber \\
&\quad+\frac{\Nf}{\pi}  \int  \frac{\dk'}{\pi} \int_0^1\dif z\;P_{\pq\pg}(z)
\Bigl[\frac{\hz_\pq(\kk')}{\qq^2 + \kk' {}^2}\;  \Sj3\big(\kk',\qq,x z;x\big)\nonumber \\
&\quad \quad-\frac{1}{\kk'{}^2} \Theta(\Lambda^2 -\kk'{}^2) V^{(0)}_{\pq}(\kk,xz) \Bigr]\nonumber \\
&\quad+\frac{\Nf}{2\pi} \int\frac{\dk'}{\pi} \int_0^1\dif z\;P_{\pq\pg}(z)
\frac{\N C_A}{(\qq-z \kk)^2}\nonumber \\
&\quad \quad\times\Bigl[ z(1-z) \frac{\qq \cdot \kk'}{\qq^2 \kk' {}^2}  \Sj3\big(\kk',\qq,xz;x\big)
\nonumber \\
&\quad \quad-\frac{1}{\kk^2}\Theta\big(\Lambda^2-(\qq-z \kk)^2\big) \Sj2(\kk,x)\Bigr]\nonumber \\
&\quad+\frac{C_A}{\pi} \int_0^1 \frac{\dif z}{(1-z)_+} \left[(1-z) P(1-z)\right]
\nonumber\\&\quad\quad\times
\int\frac{\dif\lb}{\pi\lb^2}\Biggl\{ \frac{\N C_A}{\lb^2+(\lb -\kk)^2}\nonumber \\
&\quad \quad \times\Bigl[\Sj{3}(z \kk+(1-z)\lb,(1-z)(\kk-\lb),x(1-z);x)\nonumber \\
&\quad \quad +\Sj{3}(\kk-(1-z)\lb,(1-z)\lb,x(1-z);x) \Bigr] \nonumber \\
& \quad  \quad - \Theta(\Lambda^2-\lb^2)
\left[ V^{(0)}_\pg(\kk,x)+ V^{(0)}_\pg(\kk,x z) \right] \Biggr\}\nonumber \\
&\quad + \frac{C_A}{\pi}\int\frac{\dif\kk'}{\pi}\int\dif z \left[
 P(z) \left( (1-z)\frac{\qq\cdot(\qq-z \kk)}{\qq^2 (\qq-z \kk)^2}\right.\right.\nonumber \\
&\quad \quad \times\hz_\pg(\kk') \Sj{3}(\kk',\qq,xz;x)\; +\nonumber \\
&\quad \quad -\left.\frac{1}{\kk'{}^2}\Theta(\Lambda^2-\kk'{}^2)V_\pg^{(0)}(\kk,xz) \right)
\nonumber \\
&\quad \quad \left.-\frac{1}{z\qq^2}\Theta(|\qq|-z(|\qq|+|\kk'|)) V_\pg^{(0)}(\kk',x)\right]\;.
\end{align}
Like for the quark case the gluon originated vertex
clearly depends on the jet definition and on three scales: the energy scale $s_0$ (via
the subtraction of the LL term $\propto1/z$), the factorization scale $\Lambda=\mu_F$ and
the renormalization scale $\mu$.

\section{A consistency check: jet vertex versus impact factor\labe{s:jvi}}

In this section we compare the structures of the high energy factorization formulae in the
two cases of jet cross section --- discussed in this paper and in~\cite{BaCoVa01} --- and of
inclusive partonic cross section --- presented in~\cite{Ci98,CiCo98}.  This is interesting
both for understanding the different structure of the singularities involved in the two
formulations and, on the other hand, to underline the equivalence in the LL subtraction
procedure which will allow us to use the results of two-loop calculations for the
generalization to higher orders presented in Sec.~\ref{s:vhe}.

\subsection{Jet cross section versus inclusive cross section\labe{ss:jvi}}

Let us consider the quark-hadron one-jet inclusive factorization formula
\begin{equation}\lab{incljet}
 \dJ{_{\pq H}} = \sum_{\pa}\int\dif x\,\dif\kk\,\dif\kk'\;
 h_\pq(\kk) G(xs,\kk,\kk') V_\pa(\kk',x) f_\pa(x)
\end{equation}
(cfr.~Eq.~(\ref{FF})) and compare it to the one describing the inclusive quark-parton cross
section
\begin{equation}\lab{inclpartonic}
 \dif\sigma_{\pq\pa}^{\mathrm{incl}} = \int\dif\kk\,\dif\kk'\;
 h_\pq(\kk)G(\sr,\kk,\kk')h_\pa(\kk')\;.
\end{equation}
Formally, one can obtain the former from the latter by simply replacing the partonic impact
factor of the lower incoming parton $\pa$ with the product of jet vertex and PDF --- we
refer to this product as ``jet impact factor''.
There is nevertheless an important difference between the one-loop correction to the impact
factor $h^{(1)}$ and the one to the ``jet impact factor'' 
$[V\product f]^{(1)}=V^{(1)}\product\fz+V^{(0)}\product f^{(1)}$.
In the gluon-initiated process, the jet
impact factor contains the full collinear singularities (included in $f^{(1)}$) relative to
the $\pg\to\pq$ and $\pg\to\pg$ splittings, as shown in Eq.~(\ref{singular}).  The gluon
impact factor, on the other hand, contains the full $\pg\to\pq$ collinear singularity but
only a part of the $\pg\to\pg$ one. In fact, the residue of the $1/\e$ pole of the gluon
impact factor correction $h_\pg^{(1)}$ is $\int\dif
z\;[P_{\pq\pg}(z)+P_{\pg\pg}(z)-2C_A/z]$. This is because the subtracted term $2C_A/z$ has
been taken into account in the LL contribution (see Eq.~(\ref{Bkern})).  The same is true
for quark-initiated processes, the residue of the $1/\e$ pole in the quark impact factor
correction $h_{\pq}^{(1)}$ being $\int\dif z\;[P_{\pq\pq}(z)+P_{\pg\pq}(z)-2C_F/z]$.

In the jet case, it has been possible to factorize the full collinear singularities into the
PDF because of the jet distribution, which selects only some final states out of the whole
phase space.

\begin{figure}[ht!]
  \centering \resizebox{0.3\textwidth}{!}{\includegraphics{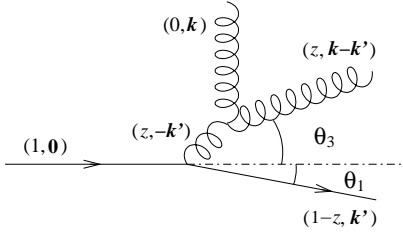}}
  \caption{\labe{f:angular}LAB frame view of the {\it quark + reggeon $\to$ quark + gluon}
    blob involved in the quark-initiated jet cross section (in particular the Feynman
    diagram responsible for the $1\parallel\pa$ collinear singularity yielding $1/\e\times
    P_{\pg\pq}$ is shown). $\theta_1$ and $\theta_3$ denotes the outgoing quark and gluon
    polar angles respectively. In parentheses close to each particles are the longitudinal
    momentum fraction (w.r.t. the incoming parton) and the transverse momentum.}
\end{figure}

How does this happen?  Let us discuss this issue in more detail for the simpler case of
quark-initiated scattering.  Fig.~\ref{f:angular} shows a possible configuration of the
quark-reggeon blob $\pq\pg^*\to\pq\pg$.  In the inclusive cross section, all configurations
contribute to the cross section. When $\theta_3<\theta_1$, the contribution is entirely
assigned to the impact factor correction.  However, when $\theta_3>\theta_1$, according to
angular ordering, the part $\propto 2C_F/z$ of the squared matrix element is interpreted as
LL term. Since the collinear region which should provide the $P_{\pg\pq}$ splitting function
times the $1/\e$ pole is just $\theta_1\simeq0$, it is clear that it is always affected by
the LL subtraction, and therefore the impact factor correction contains only the ``non
singular'' part $P_{\pg\pq}(z)-2C_F/z$ of the splitting function.

In the jet case, in the collinear region $\theta_1\simeq0$, we perform no LL subtraction,
because the gluon is constrained to be the jet, hence the $1/\e$ collinear pole multiplies
the full $P_{\pg\pq}$ splitting function (and the LO jet vertex). The LL subtraction is
performed when $\theta_3>\theta_1$ {\em and} the jet consists of only the quark. In this
case there is no collinear singularity associated with the LL subtraction.

\subsection{Relation between $h^{(1)}$ and $V^{(1)}$\labe{ss:hV}}

One could think to check the consistency of our calculation by integrating the factorization
formula of the jet observables with respect to the jet variables in order to recover the
quark-hadron inclusive cross section
\begin{equation}\lab{inclhadronic}
 \dif\sigma_{\pq H}^{\mathrm{incl}} = \sum_\pa\int\dif x\,\dif\kk\,\dif\kk'\;
 h_\pq(\kk)G(xs,\kk,\kk')h_\pa(\kk') \fz_\pa(x)\;,
\end{equation}
obtained by convoluting the quark-parton inclusive cross section~(\ref{inclpartonic})
with the bare%
 \footnote{One must use the bare PDFs because the collinear singularities are embodied in
   the partonic impact factors.}%
\ parton densities $\fz_\pa(x)$.

This is not the case, because the single-jet configurations would be
counted twice. Consider, for instance, the configuration of Fig.~\ref{f:angular}. It would
be counted once when the quark is the jet and a second time when the gluon is the jet. On
the contrary, if we replace the jet distribution $\Sj3$ with another distribution $\Sjt3$
with the property that
\begin{equation}\lab{newS}
 \int\dif J\;\Sjt3(\pp_1,\pp_3,\xi;x)=1\;,\qquad \dif J\dug\dif y_J\dif E_J\dif\phi_J\;,
\end{equation}
(i.e., a function that counts each final state once and only once),
and define a ``modified jet vertex''
$\widetilde{V}^{(1)}$, obtained from $V^{(1)}$ by replacing $\Sj3\to\Sjt3$,
then the integration over the jet variables of the ``modified one-jet inclusive cross
section'' yields the inclusive cross section~(\ref{inclhadronic}), because
\begin{equation}\lab{intVtildef1}
 \int\dif J\;\widetilde{V}\product f = h\product\fz\;.
\end{equation}
More precisely:
\begin{align}\nonumber
  &\int\dif J\;\biggl\{\widetilde{V}^{(1)}_\pa(\kk,x;J)
  +\frac1{2\pi\e}\left(\frac{\Lambda^2}{\mu^2}\right)^\e\\ \nonumber &\quad\times\int\dif
  z\; \bigl[V^{(0)}_\pa(\kk,xz;J)P_{\pa\pq}(z)+V^{(0)}_\pa(\kk,xz;J)P_{\pa\pg}(z)\bigr]\biggr\}
  \\ \lab{check} &=h^{(1)}_\pa(\kk)\;,\qquad(\pa=\pq,\pg)\;.
\end{align}

Despite of the difference between the partonic and jet impact factors, the remaining factors
in the respective factorization formulae are the same. In particular, the Green's function
at scale $|\kk||\kk'|$ in the form of Eq.~(\ref{HGH}) is identical to the one defined
in~\cite{Ci98} for the partonic case, i.e., it shares the same impact kernel $H_L$ and
$H_R$.

This is due to the method adopted for the definition of the partonic impact factor (resp.,
the jet vertex), i.e., for the separation of the central region from the fragmentation
region. In practice, whatever the selection of the final state for the observable under
study, the precise contribution of the fragmentation region of each of the incoming
particles has been defined according to the following scheme:\\
- take the expression of the squared matrix element (supplied eventually with the
distribution $\CMcal S$ providing the final state selection) valid when the gluon is in the
fragmentation region of the incoming particle;\\
- extract the central region limit from the above expression. It should be proportional to
the BFKL kernel times the LO impact factor times an enhancing factor describing the large
extension of the central region (in our variables, this enhancing factor is $1/z$), provided
the final state selection is inclusive in the central region. This is the case for the one-
and two-jet inclusive observables that we are studying;\\
- at {\em fixed quark momentum}, subtract from the fragmentation region expression the one
obtained as central region limit. This subtraction has to be performed according to the
angular ordering prescription $\theta_3>\theta_1\iff y_3>y_1$ in the soft gluon region.

This defines on one side the impact factors, and on the other side what is not impact
factor, i.e., what contributes to the Green's function. The universality of the Green's
function (at least at one-loop) should now become clearer: it is due to the same scheme of
defining the LL subtraction, yielding in all cases the BFKL kernel times the LO impact
factor. The details of the angular ordering out of the soft gluon region and the choice of
the energy scale determines the form of the impact kernel $H_{L,R}$.

\section{Factorization formula at higher orders\labe{s:vhe}}

In this section we discuss some issues related to the generalization 
of the factorization formula (\ref{ansatz}) to higher order perturbation 
theory. 

It was the main purpose of this paper (and of the previous
one~\cite{BaCoVa01}) to provide a factorization formula for the jet
production cross section at high energies, accurate in the
next-to-leading $\log s$ approximation. To this end we have studied
the jet cross section formula in next-to-leading order,
$\ord(\as^3)$. The leading logarithmic approximation (LL) retains
the coefficient of $\ln s$. What we have computed in this paper is the
next-to-leading logarithmic approximation (NLL), i.e. the terms
without the $\ln s$ enhancement. The generalization to higher order in
$\as$ therefore contains, in LL, terms of the form $\as^n
(\ln s)^{n-2}$, and in NLL contributions of the form $\as^n (\ln
s)^{n-3}$.  We are interested in summing both the LL and NLL terms to
all orders $\as$.

In generalizing our result to all orders we have to rely on Regge
factorization properties of scattering amplitudes in the high energy
limit. Regge theory states that amplitudes (both for elastic
scattering, and also for the inelastic production of particles in the
multiregge limit, when described by the exchange of Regge poles) obey
factorization properties. In QCD the gluon has been found to reggeize,
and, as it is discussed in some detail in the original BFKL
papers~\cite{BFKL76} and in the NLO calculations of the BFKL
kernel~\cite{FaLi98,CaCi98}, scattering amplitudes with color octet
exchange are expected to satisfy these factorization properties. In
QCD calculations, this leads to the `Regge ansatz' for elastic
parton-parton scattering and for inelastic gluon production in the the
multiregge limit. Strictly speaking, this ansatz is a hypothesis,
suggested by Regge theory and by the observation that the gluon
reggeizes.  The strongest support for the correctness of the
assumption comes from the result that the LL BFKL equation, derived
from the Regge ansatz, in the color octet channel satisfies the
bootstrap condition. More recently, the Regge ansatz has been used
also in NLO calculations~\cite{FaLi98,CaCi98,BaGiQi00}; the bootstrap
condition in the NLO BFKL equation was first formulated in its weak
form inside elastic scattering amplitudes and after in its strong
form, which is the one necessary for the self-consistency of the
assumption of Reggeized form of the production
amplitudes~\cite{bootstrapnloFadin,bootstrapnloBV}.  Very recently the
validity of the bootstrap condition (strong form) in the NLO BFKL
equation has been proven in~\cite{FaPa02}.

In our case, we have used the Regge ansatz for the factorizing form in (\ref{ansatz}):
in the jet production amplitude the large rapidity gap between the outgoing quark 
(at the upper end) and the jet (at the bottom) leads to the 
exchange of a reggeized gluon, and the upper and lower vertex become 
independent from each other. In the cross section formula (\ref{ansatz}) 
this factorization property leads to the impact factor and the jet vertex.
In our calculation, it has been one of our tasks to show that the jet vertex 
fits into this factorization scheme: when integrated over the jet variables, 
the square of the ``modified jet vertex'' turns into the gluon (or quark) impact 
factor which has been studied before. Without the integration over the 
jet variables we have, in principle, a new situation: 
we find that Regge factorization still works at the NLO level. 
One expects that this factorization continues 
to hold also in higher order $\as$, i.e. the impact factor at the upper 
end of the gluon and the jet vertex allow for a systematic expansion 
in powers of $\as$. Similarly, the exchanged gluon which in lowest order 
is elementary turns into a reggeized gluon with a trajectory function being 
given as a power series in $\as$.   

When generalizing to higher orders, i.e. to the production of 
more gluons, we follow the strategy pursued in the 
NLO calculations of the BFKL kernel: the Regge ansatz for multi-gluon 
production amplitudes states that vertices that are separated by large 
rapidity gaps are independent of each other, i.e. they factorize.    
Before turning to the all-order formula, it may be instructive to 
take a closer look into the next order correction to the cross section
formula, the two-loop jet cross section $\ord(\as^4)$. 

Since final states may consist of four, three, or two partons we have 
terms of order $\as^4\log^2 s$ and $\as^4\log s$ (terms without $\ln s$ 
belong to NNLL and will not be considered). Regge factorization leads us 
to expect:
\begin{align}\nonumber
 \frac1{\as^4}\dJ{^{(2)}} &= \frac12\log^2\frac{xs}{s_R}\;\hz K^{(0)} K^{(0)} V^{(0)} \fz
 \\ \nonumber
&\quad+\log\frac{xs}{s_R}\Big[h^{(1)} K^{(0)} V^{(0)} \fz
 \\ \nonumber
&\quad+\hz\big(H_L K^{(0)}+ K^{(0)} H_R+K^{(1)}\big) V^{(0)} \fz
 \\ \label{2loop}
&\quad+\hz K^{(0)} V^{(1)} \fz+\hz K^{(0)} V^{(0)} f^{(1)}\Big]\;.
\end{align}
The first term on the RHS is 
the LL term given by the iteration of the leading kernel $K^{(0)}$.
The square brackets collect the NLL contributions:  the impact factor 
correction, the
contribution from the products of impact kernel and leading kernel, 
the NLL kernel, and finally the corrections to the jet vertex and to the PDF.
\begin{figure*}[t!]
  \centering
  \resizebox{0.75\textwidth}{!}{\includegraphics{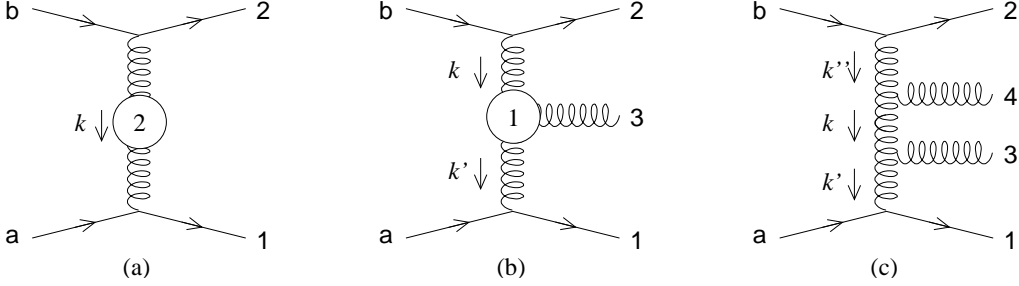}}
  \caption{\labe{f:2loop}Schematics of amplitude diagrams involved in the calculation of the
    partonic cross section at relative $\ord(\as^2)$: (a) two particle final state with
    two-loop virtual corrections; (b) three particle final state with one-loop virtual
    corrections; (c) four particle final state at tree level. The final state particles
    are ordered with increasing rapidity from the top to the bottom.}
\end{figure*}

In the second step we relate the terms of Eq.~(\ref{2loop}) to the various 
final states and kinematic regions contributing to the $\ord(\as^4)$ cross 
section. We have three kinds of final states: two-parton final states 
(with two-loop virtual corrections), three-parton final states 
(with one-loop virtual corrections) and four-parton final states at 
tree level. Because of the reggeization of the gluon, there is no one-to-one
correspondence between the number of rapidity gaps and the power of $\ln s$.
Let us go through the final states in some detail. 

Consider first the four-parton final state of Fig.~\ref{f:2loop}c. 
It can be shown that, when the two central gluons are emitted at large 
subenergies $s_{ij}\gg t_j$, i.e., final state particles are separated by 
large rapidity gaps, the cross section is described by the
iteration of two BFKL kernels (actually their real parts), 
one for each emitted gluon. Since the external partons are isolated in 
rapidity space, they are described by LO partonic
impact factors in the inclusive process, and by the LO jet vertices in the 
jet process.

NLL contributions arise from quasi-multi-regge kinematics, where one 
(and only one) pair
of particles is emitted at small subenergy (small rapidity gap). 
If the pair with small
subenergy is formed by the two central gluons 3 and 4, there is no ambiguity 
to associate the corresponding contribution to the NLL BFKL kernel $K^{(1)}$. 
This is one of the so called ``irreducible contribution'' of 
Ref.~\cite{CaCi98}, because it can directly be interpreted
as contribution to $K^{(1)}$, regardless of the definition of the 
impact factors. Note that the rapidity intervals $y_1-y_3$ and $y_4-y_2$ 
must be large in order to ensure the contribution
to be NLL. Therefore the external particles 1 and 2 are still described by 
LO impact factors or jet vertices.

When one of the central gluon approaches an external particle, 
say gluon 3 enters the fragmentation region of particle 1, $s_{13}$ and 
$y_1-y_3$ become small. In this case we
have the contribution to the jet vertex and PDF corrections. The definition 
given in the one-loop situation still works here, always because of high 
energy factorization: since
$s_{34}$ and $s_{42}$ are much larger than $s_{13}$, the emission of gluon 4 
is described by the real part of the LL BFKL kernel, independently of what 
is going on in the upper or lower vertices. Therefore, we can apply the 
machinery of the one-loop calculation factoring out the $\pb\to2+4$ 
subprocess (described by $\hz K^{(0)}_{\mathrm{real}}$, and consider only the
blob $\pa+\pg^*(k)\to1+3$ as done before.  What one obtains --- after the 
subtraction of the central region contribution of gluon 3 according to the 
angular ordering prescription --- is the sum of the one-loop (real) 
corrections to the vertex and to the PDF, the impact kernel
$H_R$, and a residual contribution that can only be assigned to the NLL 
kernel $K^{(1)}$. This
is one of the so called ``reducible'' contribution, in that it depends on the 
definition of the ``(jet) impact factors''. The key point is that also the 
reducible contributions to $K^{(1)}$ are the same as those obtained in the 
inclusive case, because they are related (just like $H_R$) to the subtraction 
of the central region from the fragmentation region, and we
have shown in Sec.~\ref{s:jvi} that the LL subtraction is defined exactly in 
the same way in the jet cross section as well as in the inclusive cross 
section.

The treatment of the phase space region where gluon 4 is close
to parton 2 is equivalent to that adopted in the inclusive process and 
provides the upper
impact factor correction $h^{(1)}$, the other impact kernel $H_L$ and 
another reducible contribution to $K^{(1)}$.

A similar analysis can be performed for the three parton final state 
emission, Fig.~\ref{f:2loop}b. When the gluon is in the central region, 
one has the product of the real part $K^{(0)}_{\mathrm{real}}$ of the kernel 
(gluon emission) and of the virtual part $K^{(0)}_{\mathrm{virt}}$ (provided by 
the leading virtual corrections), but also virtual
corrections to the partonic or jet impact factors (given by virtual 
corrections attaching only to the corresponding partonic line) and an 
irreducible contribution to $K^{(1)}$ (given by virtual corrections 
attaching to the outgoing central gluon and to the exchanged gluons).
When the emitted gluon approaches the lower partonic line, the contribution 
to the cross section can be at most NLL, and this occurs only if there are 
virtual corrections ``filling'' the large rapidity gap $y_3-y_2$ providing a 
factorized one-loop gluon trajectory (virtual part of the BFKL kernel). 
Therefore, also in this case we can factor out
the upper blob $\pb\to2$ (described by $\hz K^{(0)}_{\mathrm{virt}}$ and 
apply the procedure of separation of central region and fragmentation region 
to the lower blob $\pa+\pg^*(k)\to1+3$ as done before.  We obtain again --- 
after the subtraction of the central region contribution of gluon 3 
according to the angular ordering prescription --- the sum of the one-loop 
(real) corrections to the vertex and to the PDF, the impact kernel
$H_R$, and another reducible contribution to $K^{(1)}$.

The emitted gluon approaching the upper partonic line is treated in the 
same way of the inclusive case, giving again a correction to the upper 
partonic impact factor, the composition of the impact kernel $H_L$ with the 
virtual part of the BFKL kernel and yet another reducible contribution to 
$K^{(1)}$.

The two-parton final state contribution (see Fig.~\ref{f:2loop}a) is the 
simplest to analyze, since the jet has a trivial structure, namely it 
consists only of the lower outgoing quark, and is described by the LO jet 
vertex $V^{(0)}\propto\hz$. The comparison with the partonic case is 
therefore straightforward and we obtain: the iteration of two
gluon trajectories $\omega^{(1)}$ (to be included in $K^{(0)}K^{(0)}$), the 
product of the virtual correction to the impact factor and the gluon 
trajectory (to be included in $h^{(1)}K^{(0)}$), the product of the 
trajectory and the virtual corrections to the lower quark-reggeon vertex
(to be included in $K^{(0)} V^{(1)}$ and $K^{(0)} f^{(1)}$) and the 
two-loop gluon trajectory $\omega^{(2)}$ (the last irreducible contribution 
to $K^{(1)}$).

From Eq.~(\ref{2loop}) and from the subsequent discussion one sees that the
quantity governing the high energy behaviour of the higher order 
processes at NLL level is the BFKL kernel $\K$, which has the 
perturbative expansion
\begin{equation}\lab{svilkernel}
 \K = \as K^{(0)}+\as^2 K^{(1)}\;.
\end{equation}
Its NLL term $K^{(1)}$ has been determined for inclusive partonic scattering.  
In a perturbative expansion of the cross section, the NLL kernel $K^{(1)}$ 
enters the first time at relative order $\as^2$, i.e., at two-loop. 

The most straightforward way of generalizing to all orders in $\as$ 
is to adapt the strategy used in the NLO study of  
parton parton scattering to our case of one-jet-inclusive processes. 
The main difference between the two
cases is that in the latter the impact factor of the lower incoming parton 
is replaced by the convolution of the PDF with the jet vertex, 
as was already shown in the one-loop calculation. Therefore, we keep the 
general structure given in Eq.~(\ref{FF}), together with
the perturbative expansions~(\ref{svil}) of the energy independent factors. 
Only the Green's function $G$ needs to be extended to NLL level.

Before we write down the general all-order formula, let us return, once more, 
to the 
question of the energy scale $s_0$. For future applications it will be 
convenient to use the 
{\em symmetric Regge-type energy scale $s_R\dug|\ku||\kd|$}. As we have 
discussed before, our requirement of giving the impact factor and the PDF's 
the correct collinear properties has led to the energy scale 
~(\ref{energyscale}), and a change of the energy scale will be induced 
by the additional operators $(1+\as H_R)$, $(1+\as H_L)$.
As discussed in~\cite{Ci98}, this leads to the following form 
of the NLO BFKL Green's function for the energy scale $s_R\dug|\ku||\kd|$:
\begin{align}\nonumber
& G(xs,\ku,\kd) = \int\frac{\dif\omega}{2\pi\ui}\left(\frac{xs}{s_R}\right)^\omega
 \\ \lab{rapG}
&\qquad\times\langle\ku|(\id+\as H_L) [\omega - \K]^{-1}(\id+\as H_R)|\kd\rangle\;.
\end{align}
in terms of the impact kernels~(\ref{defHLHR}) and of the BFKL kernel~(\ref{svilkernel}).
Inserting this Green's function into~(\ref{FF}) and expanding in powers 
of $\as$ one readily reproduces~(\ref{2loop}). 

We conclude with a final formula for the Mueller-Na\-ve\-let jets:
\begin{align} \lab{finaljet}
\frac{\dif^2 \sigma}{\dif J_1\dif J_2} = \sum_{\pa,\pb} \int \dif x_1 \dif x_2
  \int \dif \kk_1 \dif \kk_2\; f_{\pa}(x_1) V_{\pa}(\kk_1,x_1) 
\nonumber \\ 
G(x_1 x_2 s,\kk_1,\kk_2) V_{\pb}(\kk_2,x_2) f_{\pb}(x_2) 
\end{align}  
where $\pa,\pb=\pq,\pg$, and the subscripts $1$ and $2$ refer to jet $1$ and $2$ in hadron 
$1$ and $2$, resp. All elements are known in LO and in NLO; in particular,
the jet vertex expressions $V_{\pg}^{(1)}$ from~(\ref{V1}), and $V_{\pq}^{(1)}$ from
(105) of~\cite{BaCoVa01}.

\section{Concluding remarks\labe{s:con}}

In this paper we have completed the analytic part of the NLO corrections 
of the jet vertex which appears in the cross section formulae for 
Mueller-Navelet jets at hadron hadron colliders and for forward jets 
in deep inelastic electron proton scattering. The final result of our 
study is summarized in eq.~(\ref{finaljet}), the analytic expression 
of the cross section for the production of Mueller-Navelet jets at the 
Tevatron or at the LHC.

Apart from the interest in performing a consistent NLO analysis of the BFKL 
Pomeron at hadron hadron colliders or in deep inelastic $ep$ scattering,
the results of our analysis may rise some theoretical interest. 
Due to the very special kinematics of the jet production processes,
the jet vertex lies at the interface between two different high energy limits,
the hard scattering regime and the Regge limit (small-$x$ limit): a priori
it was not clear whether, at NLO accuracy, it would be possible to 
separate the collinear divergences from the BFKL-type gluon production.
We have obtained an affirmative answer. 
Previous experience with partonic impact factors has served as a valuable 
guide in performing this separation inside the jet vertex.
 
What remains is the numerical evaluation of the cross section formulae,
using the results derived in this paper. As the first step,
one has to specify the jet functions $S_J^{(i)}$, i.e. we have to decide 
on a jet algorithm. In addition, experimental cuts have to be formulated.
We hope to be able to report first numerical results in the near future.   
   

\end{document}